\documentclass[sigconf]{acmart}

\usepackage{microtype}
\usepackage{graphicx}
\usepackage{booktabs} 

\usepackage{amsmath}
\usepackage{algorithm}
\usepackage{algpseudocode}
\usepackage{color}
\usepackage[T1]{fontenc}
\usepackage{soul}
\usepackage{ragged2e}
\usepackage{makecell}

\usepackage{breakurl}

\usepackage{setspace}
\usepackage{comment}
\usepackage{subfigure}
\usepackage[font=small]{caption}
\usepackage[most]{tcolorbox}
\usepackage{endnotes,xspace,fancyvrb,multirow}

\usepackage{paralist}

\newcommand{\ignore}[1]{}
\usepackage{fancyhdr}
\usepackage[normalem]{ulem}

\usepackage{tikz}

\newcommand{\sch}{Themis\xspace}

\newcommand{\schfifo}{Themis+FIFO\xspace}
\newcommand{\schscf}{Themis+SCF\xspace}
\newcommand{\schscffull}{Themis+Smallest-Chunk-First\xspace}

\newcommand{\allreduce}{All-Reduce\xspace}
\newcommand{\reducescatter}{Reduce-Scatter\xspace}
\newcommand{\allgather}{All-Gather\xspace}
\newcommand{\allreduceshort}{AR\xspace}
\newcommand{\reducescattershort}{RS\xspace}
\newcommand{\allgathershort}{AG\xspace}

\newcommand{\resnet}{ResNet-152\xspace}
\newcommand{\gnmt}{GNMT\xspace}
\newcommand{\dlrm}{DLRM\xspace}

\newcommand{\transformerlarge}{Transformer-1T\xspace}
\newcommand{\ideal}{Ideal\xspace}

\newcommand{\switchhomo}{3D-SW\_SW\_SW\_homo\xspace}
\newcommand{\switchhetero}{3D-SW\_SW\_SW\_hetero\xspace}
\newcommand{\FcRingSwitch}{3D-FC\_Ring\_SW\xspace}

\newcommand{\RingFcRingSwitch}{4D-Ring\_FC\_Ring\_SW\xspace}

\newcommand{\SwitchSwitchColored}{2D-\textcolor{teal}{SW}\_\textcolor{blue}{SW}\xspace}
\newcommand{\switchhomoColored}{3D-\textcolor{teal}{SW}\_\textcolor{blue}{SW}\_\textcolor{violet}{SW}\_homo\xspace}
\newcommand{\switchheteroColored}{3D-\textcolor{teal}{SW}\_\textcolor{blue}{SW}\_\textcolor{violet}{SW}\_hetero\xspace}
\newcommand{\FcRingSwitchColored}{3D-\textcolor{teal}{FC}\_\textcolor{blue}{Ring}\_\textcolor{violet}{SW}\xspace}
\newcommand{\RingSwitchSwitchSwitchColored}{4D-\textcolor{red}{Ring}\_\textcolor{teal}{SW}\_\textcolor{blue}{SW}\_\textcolor{violet}{SW}\xspace}
\newcommand{\RingFcRingSwitchColored}{4D-\textcolor{red}{Ring}\_\textcolor{teal}{FC}\_\textcolor{blue}{Ring}\_\textcolor{violet}{SW}\xspace}

\newcommand{\PCignore}[1]{}

\def\Snospace~{\S{}}

\newcommand{\red}[1]{\textcolor{red}{#1}}
\newcommand{\teal}[1]{\textcolor{teal}{#1}}
\newcommand{\blue}[1]{\textcolor{blue}{#1}}
\newcommand{\violet}[1]{\textcolor{violet}{#1}}

\newcommand{\insertFigureNewTop}[3]{
    \begin{figure}[t]
\setlength{\abovecaptionskip}{-1pt}
\setlength{\belowcaptionskip}{-1pt}
        \centering
        \includegraphics[width=#3\linewidth]{figs/#1.pdf}
        \caption{\small #2}
        \label{fig:#1}
    \end{figure}
}

\newcommand{\insertFigureNewSpace}[4]{
    \begin{figure}[t]
\setlength{\abovecaptionskip}{-1pt}
\setlength{\belowcaptionskip}{-1pt}
        \centering
        \includegraphics[width=#3\linewidth]{figs/#1.pdf}
        \vspace{#4mm}
        \caption{\small #2}
        \label{fig:#1}
    \end{figure}
\setlength{\textfloatsep}{4mm}
}

\newlength\mylen

\newcommand{\insertWideFigureNew}[3]{

    \begin{figure*}[h]
    \setlength{\abovecaptionskip}{-2pt}
    \setlength{\belowcaptionskip}{-4pt}
        \centering
        \includegraphics[width=#3\linewidth]{figs/#1.pdf}
	    \vspace{-1mm}
	    \bigskip
        \caption{\small #2}
        \bigskip
    	\vspace{-4mm}
        \label{fig:#1}
    \end{figure*}
}

\newcommand{\insertWideFigureNewSpace}[4]{

    \begin{figure*}[h]
    \setlength{\abovecaptionskip}{-2pt}
    \setlength{\belowcaptionskip}{-4pt}
        \centering
        \includegraphics[width=#3\linewidth]{figs/#1.pdf}
	    \vspace{#4mm}
	    \bigskip
        \caption{\small #2}
        \bigskip
    	\vspace{-4mm}
        \label{fig:#1}
    \end{figure*}
}

\newcommand{\insertWideFigureNewSpaceBottom}[5]{

    \begin{figure*}[h]
    \setlength{\abovecaptionskip}{-2pt}
    \setlength{\belowcaptionskip}{-4pt}
        \centering
        \includegraphics[width=#3\linewidth]{figs/#1.pdf}
	    \vspace{#4mm}
	    \bigskip
        \caption{\small #2}
        \bigskip
    	\vspace{#5mm}
        \label{fig:#1}
    \end{figure*}
}

\newcommand{\squishlist}{
 \begin{list}{$\bullet$}
  { \setlength{\itemsep}{0pt}
     \setlength{\parsep}{3pt}
     \setlength{\topsep}{3pt}
     \setlength{\partopsep}{0pt}
     \setlength{\leftmargin}{1.5em}
     \setlength{\labelwidth}{1em}
     \setlength{\labelsep}{0.5em} } }

\newcommand{\squishlisttwo}{
 \begin{list}{$\bullet$}
  { \setlength{\itemsep}{0pt}
     \setlength{\parsep}{0pt}
    \setlength{\topsep}{0pt}
    \setlength{\partopsep}{0pt}
    \setlength{\leftmargin}{2em}
    \setlength{\labelwidth}{1.5em}
    \setlength{\labelsep}{0.5em} } }

\newcommand{\squishend}{
  \end{list}  }

\newcommand{\betterparagraph}[1]{\textbf{#1.}}

\usepackage{xcolor}

\newcommand{\SR}[1]{#1}
\newcommand{\WW}[1]{#1}

\AtBeginDocument{%
  \providecommand\BibTeX{{%
    \normalfont B\kern-0.5em{\scshape i\kern-0.25em b}\kern-0.8em\TeX}}}

\setcopyright{acmcopyright}
\copyrightyear{2022}
\acmYear{2022}
\acmDOI{10.1145/3470496.3527382}

\acmConference[ISCA '22]{The 49th Annual International Symposium on Computer Architecture}{June 18--22, 2022}{New York City, NY}
\acmBooktitle{The 49th Annual International Symposium on Computer Architecture (ISCA '22), June 18--22, 2022, New York, NY, USA}
\acmPrice{15.00}
\acmISBN{978-1-4503-8610-4/22/06}

\begin{document}

\title{Themis: A Network Bandwidth-Aware Collective Scheduling Policy for Distributed Training of DL Models}

\author{Saeed Rashidi}
\authornote{Both authors contributed equally to this research.}
\email{saeed.rashidi@gatech.edu}
\orcid{0000-0002-6472-9920}
\affiliation{%
  \institution{Georgia Institute of Technology}
  \streetaddress{North Ave NW}
  \city{Atlanta}
  \state{Georgia}
  \country{USA}
}

\author{William Won}
\authornotemark[1]
\email{william.won@gatech.edu}
\affiliation{%
  \institution{Georgia Institute of Technology}
  \streetaddress{North Ave NW}
  \city{Atlanta}
  \state{Georgia}
  \country{USA}
}

\author{Sudarshan Srinivasan}
\email{sudarshan.srinivasan@intel.com}
\affiliation{%
  \institution{Intel}
  \city{Bangalore}
  \state{Karnataka}
  \country{India}
}

\author{Srinivas Sridharan}
\email{ssrinivas@fb.com}
\affiliation{%
  \institution{Meta}
  \city{Menlo Park}
  \state{California}
  \country{USA}
}

\author{Tushar Krishna}
\email{tushar@ece.gatech.edu}
\affiliation{%
  \institution{Georgia Institute of Technology}
  \streetaddress{North Ave NW}
  \city{Atlanta}
  \state{Georgia}
  \country{USA}
}

\renewcommand{\shortauthors}{Rashidi and Won, et al.}

\begin{abstract}

Distributed training is a solution to reduce DNN training time by splitting the task across multiple NPUs (e.g., GPU/TPU). However, distributed training adds communication overhead between the NPUs in order to synchronize the gradients and/or activation, depending on the parallelization strategy.
In next-generation platforms for training at scale,  NPUs will be connected through multi-dimensional networks 
with diverse, heterogeneous bandwidths. This work identifies a looming challenge of keeping all network dimensions busy and maximizing the network BW within the 
hybrid environment if we leverage scheduling techniques for collective communication on systems today.
We propose \sch, a novel collective scheduling scheme that dynamically schedules collectives (divided into chunks) to balance the communication loads across all dimensions, further improving the network BW utilization. Our results show that on average, \sch can improve the network BW utilization of the single \allreduce by \WW{1.72$\times$ (2.70$\times$ max)}, and improve the end-to-end training iteration performance of real workloads such as \resnet, \gnmt, \dlrm, and \transformerlarge by \WW{1.49$\times$ (2.25$\times$ max), 1.30$\times$ (1.78$\times$ max), 1.30$\times$ (1.77$\times$ max), and 1.25$\times$ (1.53$\times$ max)}, respectively.



\end{abstract}




\maketitle

\section{Introduction}\label{sec:intro}
Deep Neural Networks (DNNs) are constantly growing in demand due to their vast applicability in different areas such as computer vision \cite{ResNet,Behnam1,Behnam2}, language modeling \cite{Transformer}, and recommendation systems \cite{DLRM}. In order to improve accuracy and enable emerging applications, the general trend has been towards an increase in both model size and the training dataset \cite{OPENAI}.
This makes the task of training these DNNs extremely challenging, 
requiring days or even months if run on a single accelerator~\cite{jia2018highly, amazonHerring}. 
For example, in 2020, OpenAI set the record for training one of the largest NLP models ever, GPT-3, with 175B parameters. The training required 355 GPU years, or the equivalent of 1,000 GPUs working continuously for more than four months~\cite{fb_fsdp}. By 2021 we have already moved to training
1 Trillion parameter models as Google recently demonstrated~\cite{Transformer1T}.

\insertWideFigureNewSpaceBottom{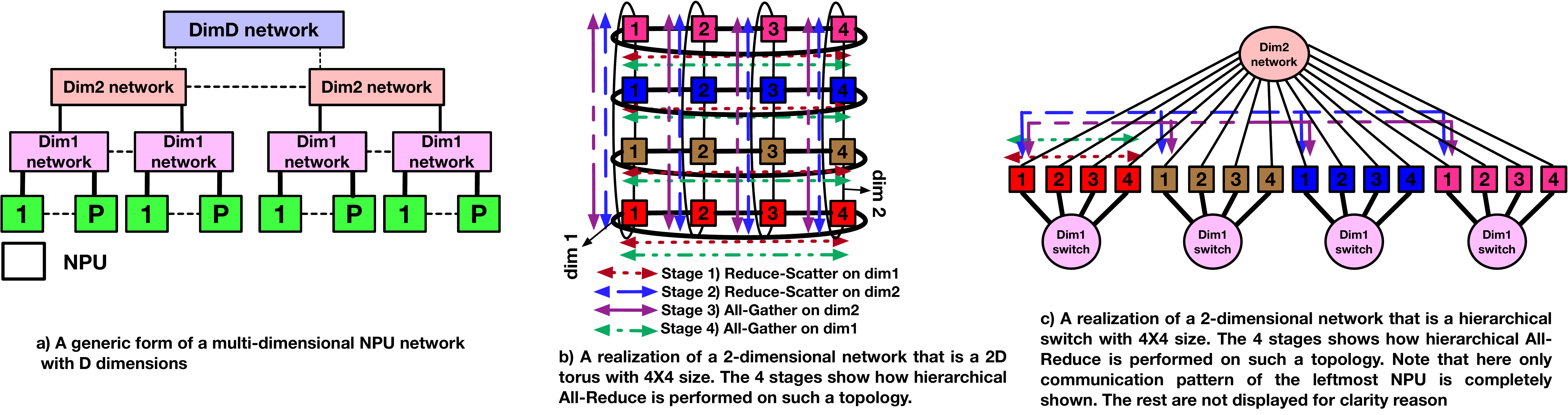}{Distributed training platform with multi-dimensional interconnection networks. Usually, the lower dimensions have higher BW, as indicated by thicker lines in the figures. There are some exceptions such as Intel Habana Gaudi \cite{intelHabana,HabanaPtP} platform, where multiple dimensions can be configured to have the same BWs.}{1}{-4}{0}


\textbf{Distributed Training Platforms.}
The challenge of training AI models has opened up a sub-field of systems research specifically aimed at designing efficient acceleration platforms for \textit{distributed training}. These platforms are built by connecting tens of high-performance accelerators (e.g., GPUs or TPUs, which we call Neural Processing Units (NPU)) together.
To leverage the compute capabilities of these platforms, the training workload (model + dataset) needs to be sharded across the accelerators via a \textit{parallelization strategy}. 
The two most popular parallelization strategies are: (i) data-parallel, where a mini-batch is split, and (ii) model parallel, 
where a model is divided across NPUs. 
Recent efforts have also looked into hybrid~\cite{DLRM} and pipelined~\cite{huang2019gpipe, harlap2018pipedream} parallelization strategies.

We identify two key trends in the \textit{network architecture} of the next-generation training platforms~\cite{intelHabana, NVIDIASuperPod, IntelExaScale, HabanaPtP, GraphCore}.

\textbf{(i) The high number of network dimensions.} 
Training platforms are built \textit{hierarchically}. 
SOTA platforms today~\cite{dgx2, googleCloudTpu} typically employ 2D network topologies --- high BW proprietary links such as XeLink \cite{IntelXeLink} or NVlink~\cite{NVlink} to interconnect NPUs on the same server followed by scaling out via NICs connected to ethernet or InfiniBand~\cite{MellanoxSHARP, NIC400G, NIC800G}. \autoref{fig:DatacenterNetwork}.a shows the abstraction of such a multi-dimensional network.  \autoref{fig:DatacenterNetwork}.b and \autoref{fig:DatacenterNetwork}.c are two realizations of such platforms, resembling today's torus-based \cite{googleCloudTpu} and switch-based \cite{dgx2} topologies, respectively.

Next-generation platforms are expected to include multiple network dimensions.
The reason for this is the growing compute and memory demand for ML models~\cite{Transformer1T}, which necessitates adding more NPUs. Adding more network dimensions is a natural way to increase scalability and overall BW per each NPU \cite{intelHabana,saeedACE}.
Further, a suite of interconnect technologies is being developed to enable scalability. For instance, there is growing interest in (i) multi-chip packaging technologies to connect several NPU dies on a package~\cite{Simba,AmdD2D400G,saeedACE,mcm-gpu}, (ii) high-bandwidth rack-scale interconnects (NVLink~\cite{NVLink3} from NVIDIA, XeLink~\cite{IntelXeLink} from Intel, Infinity Fabric~\cite{AMDInfinity} from AMD) to connect NPU packages together, and (iii) high-speed NICs and switches (e.g., Mellanox SHARP~\cite{MellanoxSHARP}) to drive high-bandwidth over Infiniband or Ethernet.

\textbf{(ii) Heterogeneity in the bandwidth of each dimension.} Generally, network bandwidth (BW) decreases as we go to the next network dimensions. However, due to recent technological advancements, the overall BW across different network dimensions can be within a comparable range.
Multi-chip packaging technology allows from 400\footnote{In this paper, all BWs are uni-directional values. The BW value gets doubled if both directions (i.e., send and receive) are considered.} Gbps \cite{Simba} to 3200 Gbps \cite{saeedACE,AmdD2D400G} BW per NPU (for NPU-to-NPU communication within a package). The rack-scale interconnect  provides up to 3600 Gbps \cite{NVLink3}. Moreover, recently 400 Gbps NICs are introduced \cite{NIC400G}, and 800 Gbps NICs will be available in the near future \cite{NIC800G}. Hence, the BW difference of the first-to-last dimension can be within 0.5--4$\times$.


These two trends lead to a challenge that this work identifies: maintaining high BW utilization across all network dimensions. This is due to the nature of 
collective communication patterns observed in distributed training. State-of-the-art collective communication (e.g., \allreduce) scheduling algorithms use hierarchical algorithms, breaking the collective into phases (e.g., \allreduce broken into Reduce-Scatter and All-Gather) and chunks from various phases of the collective moving through the network dimensions in a pipelined manner (\autoref{subsec:HierarchicalCollectives}).
Unfortunately, as we identify in this work, 
a mismatch between the chunk (scheduling unit) size and BW per dimension can lead to unbalanced pipeline stages.
This in turn means that the overall communication performance is dictated by the slowest stage, leading to network BW underutilization in other dimensions of the topology.
This mismatch arises because 
the volume of data being sent per dimension depends on the 
workload (i.e., the DNN model being trained, its parallelism strategy,  collective algorithm, and collective scheduling) while the bandwidth per dimension depends on 
system size, dollar costs, and other constraints (e.g., performance, cabling, power, etc.). 
While this is not a major problem in systems today which use few dimensions with significant BW gap across dimensions, 
this can lead to  severe network BW underutilization for next-gen platforms with the characteristics described earlier.

In this paper, we propose \sch\footnote{Themis refers to the goddess of justice, analogous to our approach that tries to uniformly balance the loads of all network dimensions.}, a novel chunk scheduling scheme that dynamically gives different chunks distinct pipeline schedules to maximize the utilization of all network dimensions. We leverage the insight that algorithmically there is no strict ordering to perform \reducescatter or \allgather stages. In other words, to perform  \reducescatter/\allgather stages a chunk may start at any network dimension and traverse dimensions in any order. The only synchronization point is that the \reducescatter stage must be completed before starting \allgather. \sch uses this fact and schedules chunks differently to balance loads of all network dimensions. \textit{\textbf{Having intelligent schedulers like \sch is a key enabler for building next-gen platforms, letting system designers design the network with respect to their metrics (e.g, cost, performance), without concerning how to efficiently utilize the network BW.}}

In short, we make the following contributions:
\vspace{-0.5mm}
\squishlist
    \item This is the first work, to the best of our knowledge, exploring the problem of multi-rail collective-communication scheduling at scale (1024 NPUs in this case) over next-gen hierarchical topologies.
    \item This is the first work to identify the problem of unbalanced stage latencies in multi-rail collective scheduling algorithms 
    and show why this leads to BW underutilization for next-gen platforms. 
    \item We propose \sch, a novel chunk scheduling scheme for multi-dimensional networks that dynamically schedules the chunks to maximize the utilization of each dimension. \sch is the first method, to the best of our knowledge, that proposes \textbf{\textit{dynamic}} scheduling for different chunks for maximum BW utilization.
    \item Our results (see \autoref{sec:methodology} for methodology) show that, on average, \sch achieves \WW{1.72$\times$} All-Reduce time speedup and \WW{95.14\%} BW utilization. This improves end-to-end training latency for \resnet, \gnmt, \dlrm, and \transformerlarge by \WW{1.49$\times$ (2.25$\times$ max), 1.30$\times$ (1.78$\times$ max), 1.30$\times$ (1.77$\times$ max), and 1.25$\times$ (1.53$\times$ max)}, respectively.
    \item 
    Using our analysis of efficient scheduling, we formulate different scenarios regarding the network BW distribution and give insights to the network designers for efficient BW distribution for the multi-dimensional networks tailored for large-scale training.
\squishend


\vspace{-1mm}
\section{Background}\label{sec:background}

\insertFigureNewSpace{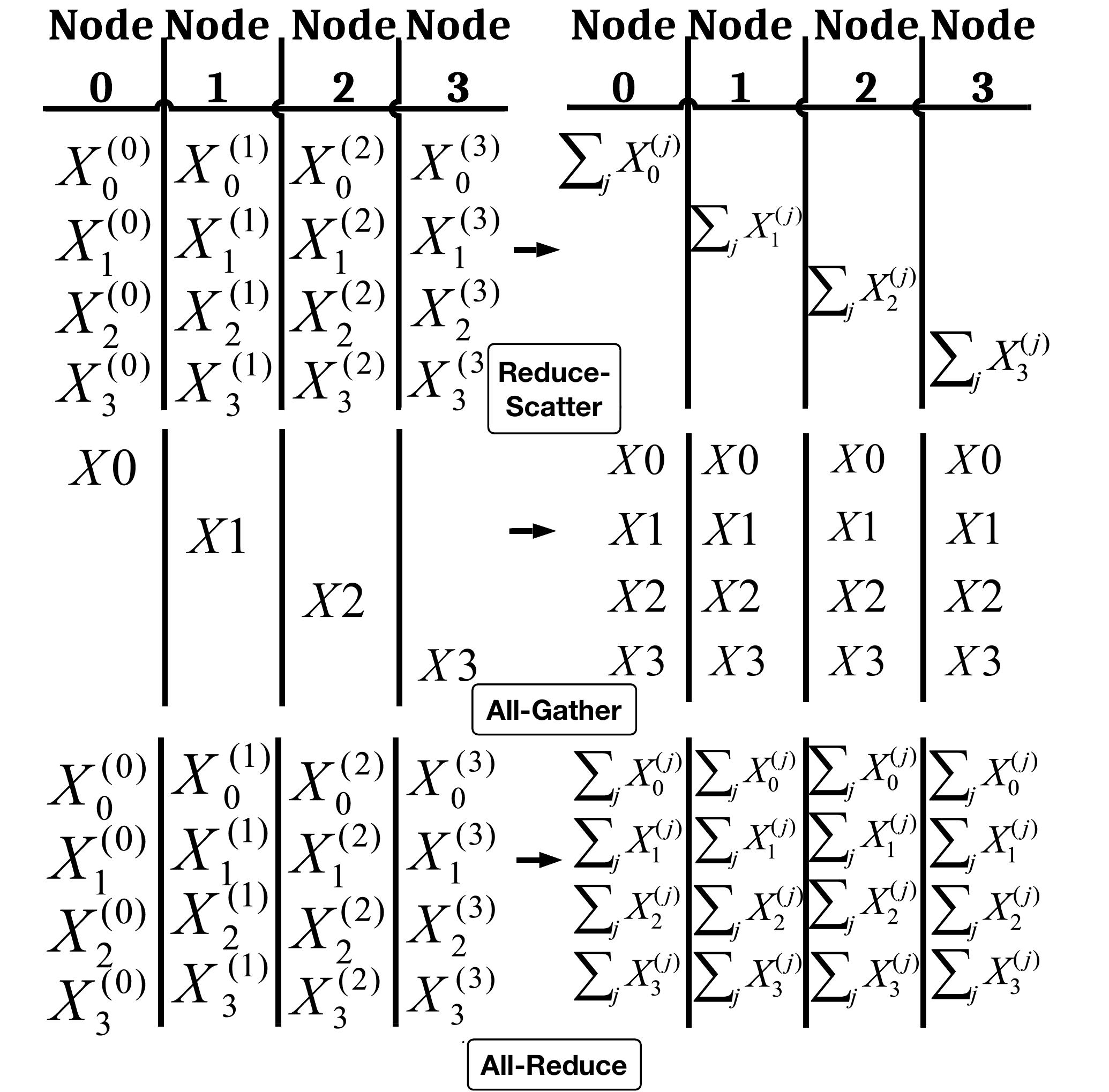}{The Mathematical Implications of the \reducescatter, \allgather, and \allreduce patterns executing on four NPUs. The left part shows the initial data on each communicating NPU before the collective operation. The right part shows data residing on each NPU after the completion of the collective operation.}{0.9}{1.5}

\insertWideFigureNew{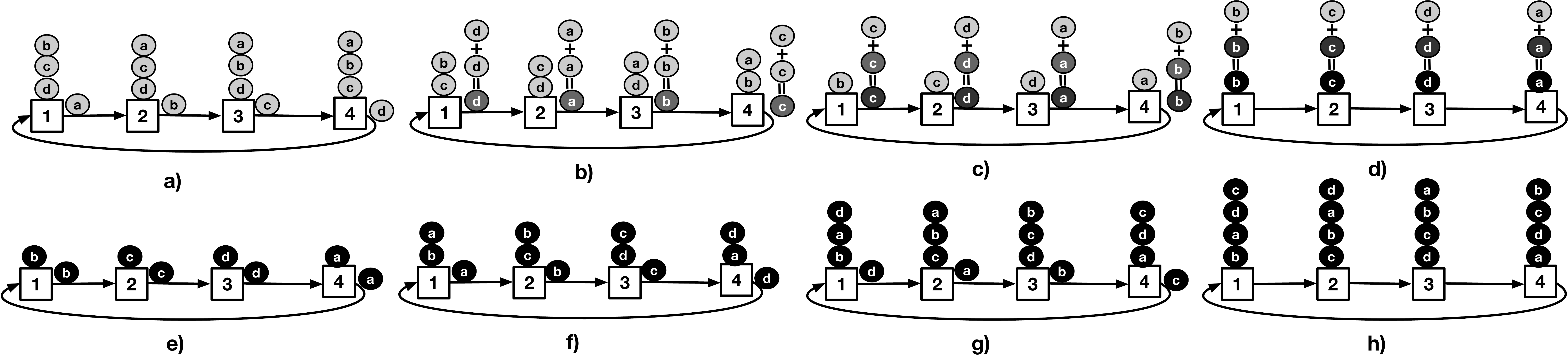}{An example of the ring \allreduce algorithm to perform the \allreduce pattern. Steps a-d perform \reducescatter pattern and steps e-g perform \allgather pattern. Step h shows the final result.}{0.9}

\subsection{Collective Communication Patterns}\label{subsec:CollectivePattern}


Communication is the inevitable overhead to pay in distributed training workloads. The exact communication patterns each training workload requires depend on the parallelization strategy, and also the communication mechanism (i.e. parameter server vs. explicit NPU-to-NPU). When using explicit NPU-to-NPU communication mechanisms, \allreduce is the most dominant pattern observed in distributed training \cite{NVidiaSwitch}\footnote{For example, in the case of a data-parallel parallelization strategy, each NPU works on a subset of the global mini-batch in each iteration, thus, their calculated weight gradients must be globally reduced (i.e. \allreduce) before updating the weights and starting the new training iteration.}. 

\allreduceshort can be broken into a \reducescatter (\reducescattershort) followed by an \allgather (\allgathershort) communication pattern. \autoref{fig:CollectiveOperations} shows the mathematical implications of these patterns performed on four NPUs. \reducescattershort performs reduction among initial data such that at the end, each NPU holds a portion of the globally reduced data. \allgathershort, on the other hand, broadcasts data residing on each NPU to all other NPUs. Therefore, it is clear that when performing \reducescattershort/\allgathershort on $P$ participating NPUs, the data size residing on each NPU shrinks/multiplies by $P\times$.



\begin{table}[!t]
\centering
\caption{Topology options per dimension and corresponding contention-free topology-aware All-Reduce algorithms.}
\label{table:TopologyAwareCollectives}
\vskip -0.1in
\begin{tabular}{|c|c|}
\hline
  \textbf{Topology} & \textbf{Topology-aware Collective} \\ \hline
     Ring & Ring \cite{collective1}\\ \hline
  FullyConnected & Direct \cite{collective1}\\ \hline
  Switch & HalvingDoubling \cite{EFLOPS}\\ \hline
\end{tabular}%
\end{table}
\setlength{\textfloatsep}{4mm}

\vspace{-1mm}
\subsection{Basic Collective Communication Algorithms}\label{subsec:BasicCollectiveAlgorithm}



Each of the collective communication patterns described in \autoref{subsec:CollectivePattern} can be performed through different \textbf{collective communication algorithms}. For example, tree-based \cite{collective3}, ring-based \cite{collective2}, and halving-doubling \cite{EFLOPS} algorithms are proposed to realize \allreduceshort pattern and are implemented in communication frameworks such as Intel oneCCL \cite{oneccl} or NVIDIA NCCL \cite{nccl}. \autoref{fig:RingAllReduce} shows an example of the ring-based \allreduceshort algorithm running on four NPUs.

The optimal collective algorithm is usually dependent on the physical topology and communication size \cite{collective1}. 
For example, ring-based collective algorithms are a natural fit for NPUs connected via a physical ring, as it leads to zero contention.
\autoref{table:TopologyAwareCollectives} presents some \textit{topology-aware collective} algorithms, which are typically chosen dynamically by communication libraries~\cite{oneccl,nccl} depending on the underlying topology.
Such basic collective algorithms provide a basis to design more complex and tuned algorithms that are optimized for multi-dimensional network topologies, as we describe next.


\vspace{-1mm}
\subsection{Multi-Rail Hierarchical Collective Comm. Algorithms}\label{subsec:HierarchicalCollectives}

As stated in \autoref{subsec:BasicCollectiveAlgorithm}, the optimal collective algorithm depends on the physical topology. Hence, the basic algorithms are not a good fit when having multi-dimensional physical networks with variable BW and latencies in each dimension. This is because the collective algorithms are inherently synchronous and in this case, the links with the least BW will become the bottleneck, making other high-BW links underutilized. To cope with this, recent works propose multi-rail hierarchical algorithms to exploit different dimensions' BW and latency \cite{blueconnect,BLink}. Suppose the 
topology has $D$ dimensions as shown in \autoref{fig:DatacenterNetwork}.a, the \allreduceshort algorithm breaks into the following 2$\times$D pipeline stage \textbf{scheduling}: 

\squishlist
    \item A sequence of \reducescattershort stages starting from dim1 and ending at dim$D$ ($D$ stages in total). After these stages, data is globally Reduce-Scattered across all NPUs.
    \item Next, a sequence of \allgathershort stages are performed in the reverse order ($D$ stages in total); starting from dim$D$ and ending at dim1.
\squishend

The above order is the \textbf{baseline collective scheduling}, used by SOTA collective libraries today~\cite{BLink, blueconnect}\footnote{If the requested collective is only \reducescattershort/\allgathershort, only the first/second half of \allreduceshort stages described above, that is \reducescattershort/\allgathershort stages, are performed.}. The main reason for such hierarchical phases is to reduce traffic as the collective goes to the next dimension, which usually has lower BW compared to the previous dimension. The \reducescattershort/\allgathershort algorithm for each stage is a basic topology-aware collective (\autoref{subsec:BasicCollectiveAlgorithm}) and is independently selected by the collective  scheduler~\cite{blueconnect, oneccl}.
For example, a topology with rings in the first and switches in the second dimension may run a series of \reducescattershort/\allgathershort stages using ring-based and halving-doubling algorithms, respectively. 



\insertWideFigureNewSpaceBottom{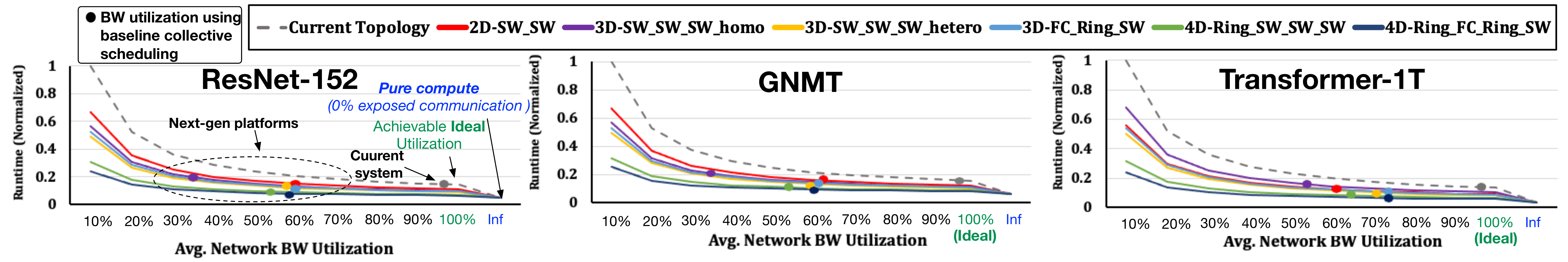}{Normalized runtime vs. average BW utilization for three different-sized DNN training workloads running on a current topology, a 2-dimensional topology with the characteristics similar to \cite{dgx2}, and six different next-gen topologies (please see \autoref{sec:methodology} for full description of workloads and topologies). The DNNs are: (i) \resnet, (ii) \gnmt, and (iii) \transformerlarge (1 Trillion params). \textit{\textbf{The avg. BW utilization is the weighted average of BW utilization across all dimensions of the network, with respect to the BW budget of each dimension (i.e., dimensions with higher BW get higher weight). It is obtained only during the time when there are communication operations issued by the workload (excluding the times when there is no pending communication operation)}}. All runtimes are normalized to the slowest topology (i.e., current topology) runtime at 10\% BW utilization. The bold dots show the BW utilization when the multi-rail hierarchical algorithm with baseline scheduling (discussed in \autoref{subsec:HierarchicalCollectives}) is used. 
\emph{Inf} (i.e., infinite) BW is when communication overhead is 0\% and runtime stems from compute only. The  \emph{ideal} is the achievable runtime if the network BW across all dimensions is fully utilized. For a fair comparison of network performance, 
we assume all systems use the same compute model 
(\autoref{sec:methodology}).}{1}{-4}{1}

\autoref{fig:DatacenterNetwork}.b and \autoref{fig:DatacenterNetwork}.c show two examples of how this \allreduceshort algorithm is applied on a 2-dimensional network. In both examples, the first dimension comprises the NPUs with the same color, meaning that the peer NPUs for the communication is the NPUs with the same color. The second dimension is shown based on the NPUs with the same number. Throughout this paper, we use the notation $P_{1}\times P_{2}\times ....\times P_{D}$ to refer to the size of a multi-dimensional network where $P_{i}$ is a number referring to the size of peer NPUs participating in the communication on the i'th dimension. For example, the size of both \autoref{fig:DatacenterNetwork}.b and \autoref{fig:DatacenterNetwork}.c is $4\times4$.

\textbf{Chunks.} Communication data is usually broken into multiple chunks~\cite{GC3,TACCL,PLink}  and then these chunks are fed into this 2$\times$D-stage pipeline to keep all dimensions busy. A chunk is a portion of data to participate in the collective, and the collective algorithm can work on each chunk independently. For example, a 256MB \allreduceshort can be broken into four independent chunks of 64MB All-Reduces. In this paper, we assume the size of each chunk in each stage to be the size of the corresponding chunk data residing on each NPU \textbf{before} the stage begins. Similar to the explanation of \autoref{subsec:CollectivePattern}, each chunk size changes after each stage of \reducescattershort/\allgathershort.

\insertWideFigureNew{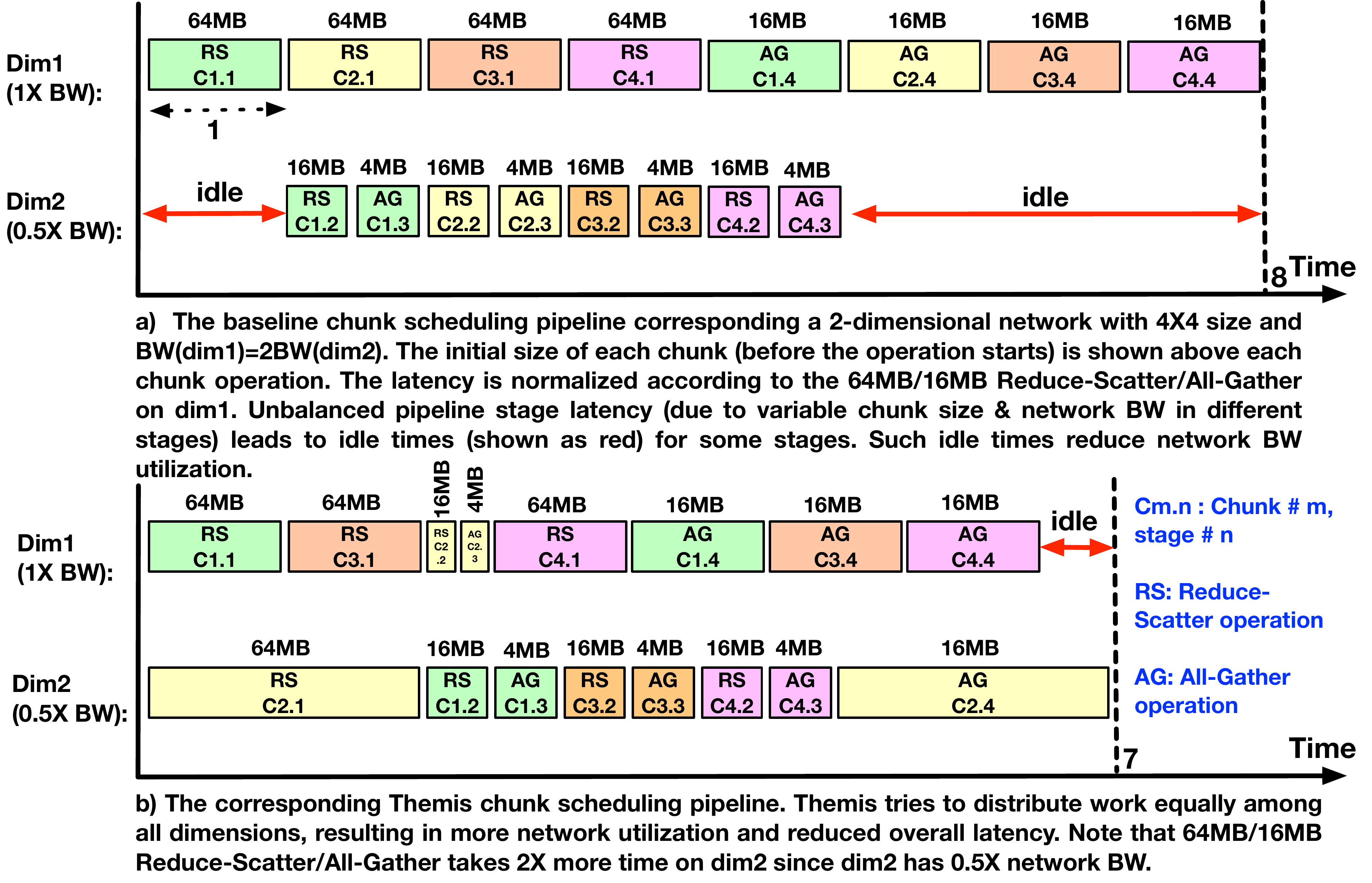}{The execution of a 256MB \allreduce running on a 4$\times$4 2-dimensional network where BW(dim1)=2BW(dim2). The \allreduce is broken into 4$\times$64MB chunks.}{0.8}
\vspace{-1mm}
\section{Motivation - Network BW underutilization}\label{sec:motivation}
\vspace{-1mm}

As discussed in \autoref{subsec:HierarchicalCollectives}, hierarchical collectives are the SOTA method for the multi-dimensional networks with variable BW. 
However, we identify that reaching the maximum possible network utilization is quite challenging for next-gen platforms.

\textbf{Definition: Average BW Utilization.} We define avg. BW utilization is the weighted average of BW utilization across all dimensions of the network, with respect to the BW budget of each dimension (i.e., dimensions with higher BW get higher weight). In the case of real workloads, it is estimated only during the time when there are communication operations issued by the workload (excluding the times when there is no pending communication operation due to compute/memory operations~\cite{saeedACE}).

\subsection{Next-Gen Distributed Training Platforms}\label{subsec:modelingSystems}

Today's high-end training platforms typically use 2-dimensional topologies \cite{dgx2,FBDLRMPlatform} -- one to connect several NPUs within the same server node together using high-speed proprietary links~\cite{IntelXeLink, NVLink3} -- followed by node-to-node communication via Network Interface Cards (NICs).
We model such 2D networks but with high bandwidth across both dimensions as a result of recent technology advancements (such as Intel's XeLink~\cite{IntelXeLink}, NVIDIA's NVlink~\cite{NVLink3}) and high-speed NICs~\cite{NIC400G,NIC800G} on next-gen platforms.

However, as workload model sizes increase \cite{Transformer}, the need for more NPUs and higher communication bandwidth increases. There is thus a growing industry trend \cite{NvidiaH100, HabanaPtP, IntelExaScale, NVIDIASuperPod, GraphCore, IntelPonteVecchio} to increase the number of dimensions \textit{before getting to the NIC} to reduce the NIC traffic. We model multiple such futuristic 3-dimensional training platforms in this work as described in \autoref{table:topology_description}. The first dimension (\textit{dim1}) represents the intra-node dimension where NPUs on the same server node are connected through a high-BW rack-scale fabric. Several nodes are then connected by allocating a portion of the rack-scale fabric to create \emph{pod} (\textit{dim2}) \cite{NvidiaH100, HabanaPtP}, again, using high-BW dedicated links\footnote{One instance is the Intel Gaudi platform \cite{HabanaPtP}, where each NPU has multiple rack-scale links that can be split for intra-node and inter-node (pod) connectivity.}. In the third dimension, NPUs within dim2 are connected to the \textit{dim3} switches through NICs. 4-dimensional topologies extend the 3D topologies by adding Multi-Chip packaging~\cite{mcm-gpu,Simba} as the first dimension to incorporate multiple NPUs within a package~\cite{mcm-gpu,AMDInfinity,saeedACE}.
\autoref{sec:methodology} provides more details of our methodology for modeling these platforms.

\vspace{-1mm}
\subsection{Quantifying Network BW Underutilization}

\autoref{fig:motivation_util} shows the overall (normalized) training time reduction as the avg. network BW utilization increases for three different DNNs with a high ratio of communication to compute. The modeled platforms are a suite of 2D, 3D, and 4D topologies as explained in \autoref{subsec:modelingSystems} and \autoref{table:topology_description}. Each line in \autoref{fig:motivation_util} shows the normalized runtime (y-axis)  for different BW utilization (x-axis) of a specific topology. 
The runtime curves are relatively similar across the three training workloads. This is because these workloads are communication bound, hence, their runtime is mainly dictated by the underlying network performance.

As \autoref{fig:motivation_util} shows, adding more network dimensions usually results in lower end-to-end training runtime, due to increased network BW per NPU. This is the motivation for the next-generation training platforms to add more network dimensions. However, the overall network BW utilization starts to drop as we add more dimensions and/or have high BW across all network dimensions, as we discuss next.
%
%


We observe from \autoref{fig:motivation_util} that a current topology can achieve 97.7\% BW utilization with the baseline collective scheduling policy as discussed in \autoref{subsec:HierarchicalCollectives}. 
This is primarily due to the huge BW difference between dim1 and dim2 (i.e. 1200 Gbps vs. 100 Gbps), making underutilization of dim2 play an insignificant role in overall performance. However, as stated earlier, next-gen platforms have high BW across dimensions, as well as having more network dimensions.

\autoref{fig:motivation_util} also shows how baseline collective communication scheduling fails to efficiently utilize the available BW on these next-gen topologies, reaching the average BW utilization of 59.7\% (35.1\% min), when averaging across all the workloads and topologies. To obtain linear (perfect) speedup as we scale the number of NPUs for training, the communication overhead should remain 0\% (the Inf BW case in \autoref{fig:motivation_util}). However, this is not feasible due to finite network BW resources (technology constrained in each dimension). Hence, for a given topology, the maximum achievable speedup is when BW utilization is 100\% (``Ideal" in \autoref{fig:motivation_util}), and any network underutilization diminishes the benefits of scaling. For the next-gen topologies, if the ideal utilization (100\%) can be achieved, the training performance on average can be improved by 1.54$\times$ (2.34$\times$ max), 1.32$\times$ (1.81$\times$ max), and 1.26$\times$ (1.54$\times$ max) over the baseline for \resnet, \gnmt, and \transformerlarge, respectively.


\vspace{-1mm}
\subsection{Understanding Network BW Underutilization}


To illustrate the problem, \autoref{fig:ThemisTimeDiagram}.a shows how a 256MB baseline hierarchical \allreduceshort is performed on a 2-dimensional network with the second dimension having half BW of the first dimension. The collective is broken into 4$\times$64MB chunks.  There are 4 pipeline stages for performing hierarchical \allreduceshort on this network:
\textcircled{1} \reducescattershort on dim1, \textcircled{2} \reducescattershort on dim2, \textcircled{3} \allgathershort on dim2, \textcircled{4} \allgathershort on dim1.



A 64MB chunk size will be shrunk by 4$\times$ when entering stage 2, meaning that \reducescattershort on the stage injects $\frac{1}{4}\times$ data to the dim2 compared to the stage 1 injecting to dim1. However, dim2 has $\frac{1}{2}\times$ BW compared to the dim1. 
If we assume the 64MB \reducescattershort (or 16MB \allgathershort) takes 1 unit of time when running on dim1, then the latency of that chunk for the stage 2 is:$\frac{0.25 Data}{0.5 BW}=0.5$. Therefore, stage 2 is performing 2$\times$ faster than stage 1. Stage 3 injects the same amount of data as stage 2 and operates on the same dimension, hence its latency is similar to stage 2. Using the same argument, stage 4 has the same latency as stage 1.
The faster processing of stage 2 and stage 3 means they are underutilized many times. This indicates that their corresponding network dimension (i.e. dim2) is underutilized as shown in \autoref{fig:ThemisTimeDiagram}.a.

We note that the only place where the baseline algorithm can reach near 100$\%$ utilization is when the BW reduction ratio in the next dimension is proportional to the size of the current dimension. Again, consider the $4\times4$ system size. For the baseline system to be efficient, the BW(dim1)=4$\times$BW(dim2) because dim1 shrinks the chunk size by $4\times$. It is only in this case that stage latencies will be equal using the baseline algorithm. Any excess BW of dim2 beyond this point will be wasted, as when we show in \autoref{fig:ThemisTimeDiagram}.a where BW(dim1)=2BW(dim2). 

In general, this concept can be generalized to any D-dimensional network of size $P_{1}\times P_{2}\times ....\times P_{D}$. For the baseline to be efficient, we must have:

$\text{BW(dim1)}=P_{1}\times \text{BW(dim2)}=P_{1}\times P_{2}\times \text{BW(dim3)}=...=P_{1}\times P_{2}\times ....\times P_{D-1}\times \text{BW(dimD)}$.

However, this creates an unpleasant requirement for the network as a result of the poor scheduling of the baseline algorithm. \SR{If we plug the above formula for the current platform (discussed in \autoref{fig:motivation_util}), we find out that all dim1 BW (1200 Gbps) and 75 Gbps (out of 100 Gbps) of dim2 are utilized using baseline collective scheduling, justifying its high BW utilization in \autoref{fig:motivation_util}.} But the underlying network dimensions can have more BW available in next-gen systems and the algorithm must be able to utilize excess BW provided by the network dimensions.

Note that in the above example, our analysis was based on the assumption that the network BW is the primary factor that determines communication latency (which is true for large collectives). However, the concept of unbalanced stage latencies remains true even if we take into account other factors (e.g. link latency) as we show in \autoref{sec:design} and \autoref{sec:results}. \textit{We also wish to emphasize that the underutilization we are referring to is a fundamental challenge due to chunk size and bandwidth mismatch (as our example indicated) and not due to any network stalls because of compute/memory bottlenecks that limit network performance~\cite{saeedACE, NVidiaSwitch} (which may further exacerbate this issue) but are not the focus of this work.}

\insertFigureNewTop{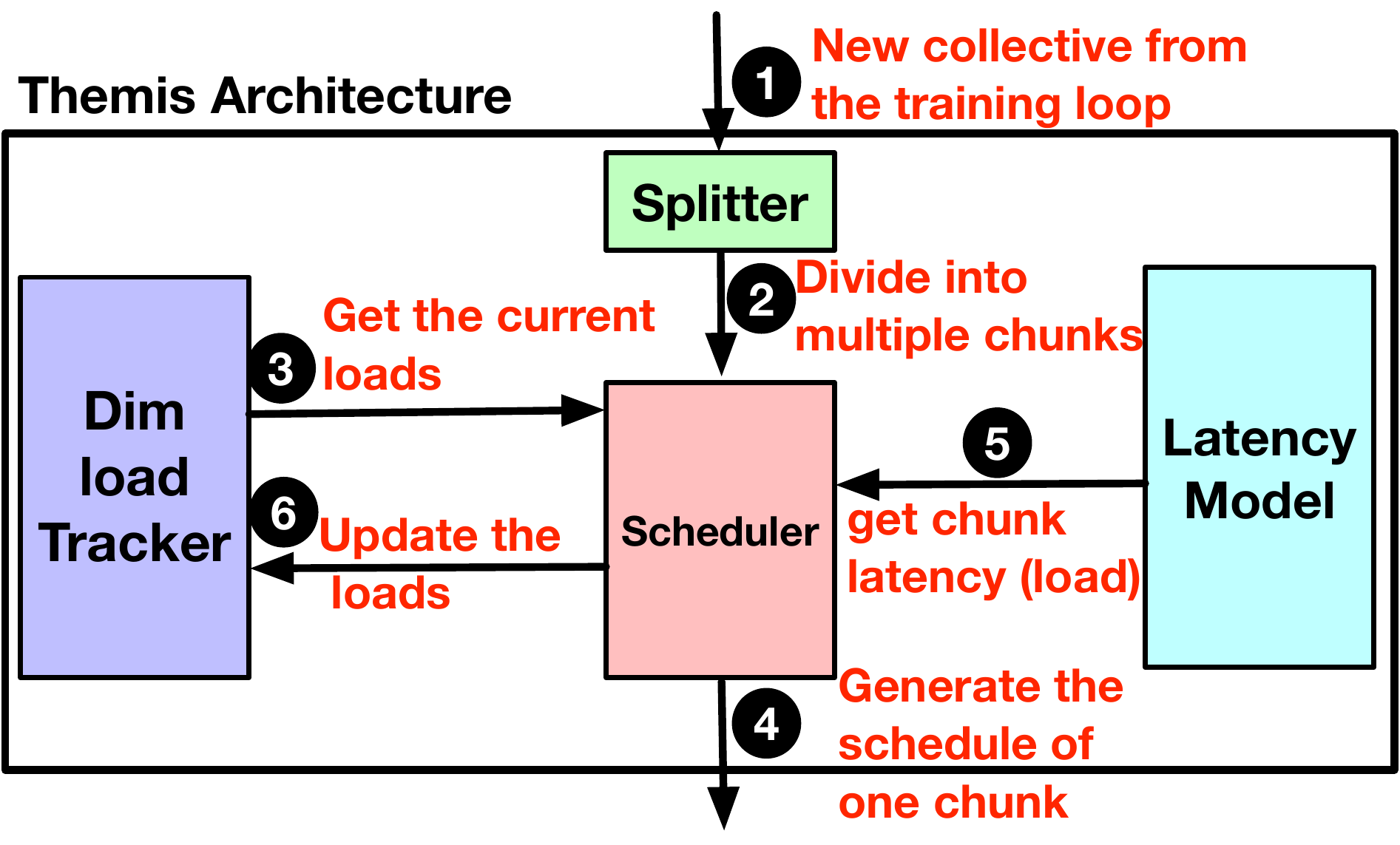}{An overview of \sch components. 1) A collective operation is requested from the upper layer training workload. 2) The collective is split into multiple equal size chunks. Steps 3--6 are performed on an individual chunk basis. 3) the current load of network dimensions (in terms of total communication latency) is retrieved from the \texttt{Dim Load Tracker}. 4) \texttt{Scheduler} sorts the dimension loads in ascending/descending orders and the sorted list order is the schedule for the current chunk through the \texttt{Latency Model}. 5) Based on the schedule generated, \texttt{Scheduler} finds out the latency of the new schedule for each dimension. 6) \texttt{Scheduler} updates the total loads of each dimension to take into account the load of the new chunk.}{0.8}
            

\begin{algorithm}[h]
  \caption{\sch Algorithm}\label{alg:Themis}
  \begin{flushleft}
  \textbf{Inputs: } CollectiveType ($CT$),  CollectiveSize ($CS$), ChunksPerCollective ($CPC$), TotalNPUs ($P$)\\
  \textbf{Output:} A 2D list $Schedule[][]$ where $Schedule[i][]$ gives the order of dimensions the i'th chunk should traverse for the collective.
  \end{flushleft}
  \begin{algorithmic}[1]
    \Procedure{Schedule\_Collective}{$CT, CS, CPC$}
      \State DimLoadTracker.reset($CT$)
      \State ChunkSize=$CS/CPC$
      \State i=0
      \For{\texttt{i++ < $CPC$}}
        \If {$CT== \text{All-Reduce}$}
            \State RS\_Sch=SCHEDULER.SCHEDULE($RS,ChunkSize$)
            \State AG\_Sch=reverseOrder($RS\_Sch$)
            \State $Schedule[i][]$=concatenate(RS\_Sch, AG\_Sch)
        \Else
            \State $Schedule[i][]$=SCHEDULER.SCHEDULE(
            \State \hspace{120pt}$CT$, ChunkSize)
        \EndIf
      \EndFor
      \State \textbf{return} $Schedule$
    \EndProcedure
    \Procedure{Scheduler.Schedule}{$CT, ChunkSize$}\Comment{\textit{Schedules a chunk}}
        \State loads=DimLoadTracker.getLoads()
        \If {{\small loads.max\_dim\_load - loads.min\_dim\_load $<$ Threshold}}
            \State schedule=getBaselineScheduling($CT$)
        \Else
            \If {$CT== \text{Reduce-Scatter}$}
                \State schedule=getIndexOfSortedList($loads,ascending$)
            \ElsIf{$CT== \text{All-Gather}$}
                \State schedule=getIndexOfSortedList($loads,descending$)
            \EndIf
        \EndIf
        \State newLoad=LatencyModel.calcLoads(
        \State \hspace{110pt}$chunkSize,schedule,CT$)
        \State DimLoadTracker.update(newLoad)
        \State \textbf{return} schedule
    \EndProcedure
  \end{algorithmic}
\end{algorithm}
\insertWideFigureNewSpaceBottom{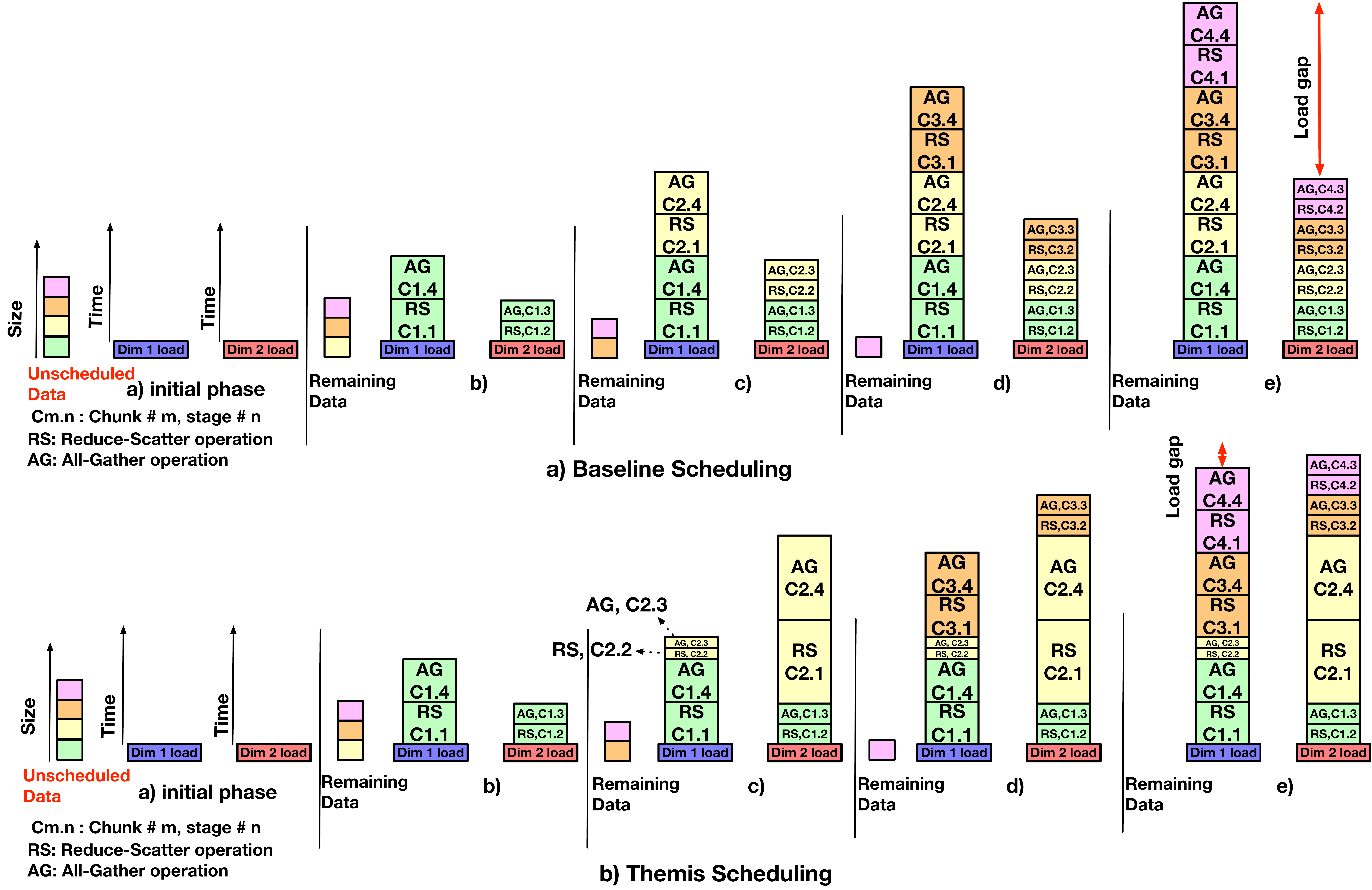}{An example of Baseline Scheduling vs. \sch Scheduling corresponding to the chunk scheduling problem of \autoref{fig:ThemisTimeDiagram}. Each of the steps b-d shows scheduling for one chunk. Dim load shows the total communication time of each dimension. As shown in the figure, baseline scheduling always uses a constant schedule for all chunks, resulting in the underutilization of abundant BW provided by dim2. However, \sch uses a greedy scheme to schedule new chunks in a way that puts more load on the dimensions with lower loads. In \sch: step b) all dim loads are zero, thus, the first chunk schedule is similar to the baseline. In step c) dim2 has a lower load, hence the chunk scheduling starts from dim2 to fill the gap with dim1. In step d) and step e) Chunk schedule starts from dim1 to fill the gap with the overloaded dim2.}{0.93}{0}{0}

\section{\sch}\label{sec:design}


In this section, we present \sch that performs dynamic and distinct scheduling for each chunk to balance the loads across different network dimensions. \sch is specifically designed to maximize the multi-dimensional network BW for \textbf{\allreduce (\allreduceshort), \reducescatter (\reducescattershort)}, and \textbf{\allgather (\allgathershort)}. 

\subsection{Themis Intuition and Overview}\label{subsec:Overview}

 \sch has its roots in two main observations:

\textbf{Observation 1.} From the algorithm's \textit{correctness} point of view, there is no restriction on how each chunk should traverse the \reducescattershort/\allgathershort stages. For example, in the case of \allreduceshort on a 2D network explained in \autoref{sec:motivation}, the \reducescattershort stage on the second dimension can precede the \reducescattershort on the first stage. Similar ordering independence is true for \allgathershort stages. Furthermore, the ordering of \reducescattershort stages can be different than \allgathershort stages. The only synchronization point is that the  \reducescattershort stages should be finished before starting the \allgathershort stages. Thus, the following 4 different schedules are all possible for the \allreduceshort collective on a 2D topology: 

\begin{enumerate}
    \item \textbf{(i)} \reducescattershort on dim1, \textbf{(ii)} \reducescattershort on dim2, \textbf{(iii)} \allgathershort on dim2, \textbf{(iv)} \allgathershort on dim1 (this is the baseline scheduling).
    \item \textbf{(i)} \reducescattershort on dim2, \textbf{(ii)} \reducescattershort on dim1, \textbf{(iii)} \allgathershort on dim2, \textbf{(iv)} \allgathershort on dim1.
    \item \textbf{(i)} \reducescattershort on dim1, \textbf{(ii)} \reducescattershort on dim2, \textbf{(iii)} \allgathershort on dim1, \textbf{(iv)} \allgathershort on dim2.
    \item \textbf{(i)} \reducescattershort on dim2, \textbf{(ii)} \reducescattershort on dim1, \textbf{(iii)} \allgathershort on dim1, \textbf{(iv)} \allgathershort on dim2.
\end{enumerate}
In general, for any D-dimensional network, there are $D!\times D!$ valid ways to schedule an \allreduceshort data chunk ($D!$ for \reducescattershort/\allgathershort only).  

\textbf{Observation 2.} Different chunks can have different schedules. Hence, if we divide a collective into C chunks, the space of all possible schedules for all chunks on an N-dimensional network is $(D!\times D!)^C$ for \allreduceshort ($D!^C$ for \reducescattershort/\allgathershort), indicating the exponential growth as the network dimensions and the number of chunks increase.  
\textit{Together, observation 1 and observation 2 motivate the \sch idea which is to independently schedule chunks across network dimensions based on the available bandwidth, rather than following a strict order like previous works~\cite{nccl,blueconnect,BLink}. For each dimension, each chunk uses the topology-aware collective algorithm for that dimension (\autoref{subsec:HierarchicalCollectives}), just like the baseline hierarchical algorithm}.


\textbf{Themis Overview.}
\autoref{fig:ThemisStructure} shows the general overview of \sch. \emph{Splitter} component simply divides the collective into multiple equally-sized chunks. \texttt{Dim Load Tracker} maintains the load of each network dimension in terms of the total communication time of the chunks when executing on that dimension. The \texttt{Latency Model} predicts the \reducescattershort/\allgathershort communication time for a given chunk size running on any network dimension. Finally, \texttt{Scheduler} generates a dynamic schedule for each chunk based on the information provided by the \texttt{Dim Load Tracker} and \texttt{Latency Model}. \autoref{fig:ThemisStructure} also shows the series of steps when a new collective communication is issued from the training workload layer. We describe its workflow next.


\subsection{\sch Algorithm}\label{subsec:Algorithm}
In \autoref{subsec:Overview} we showed how the space of available schedules grows exponentially. Therefore, it is not practical to search all possible schedules, even for modest network and chunks granularity\footnote{For example, when $D$=3, and $C$=8, all the space of all All-Reduce schedules is $(3!*3!)^8=2821109907456.$}. Instead, \sch is a type of greedy algorithm that tries to schedule new chunks in a way that puts more load (in terms of communication time) on the dimension with an already lower load. \sch leverages the two observations explained in \autoref{subsec:Overview} for more flexible scheduling. Algorithm \autoref{alg:Themis} shows the pseudo-code for \sch. The \emph{SCHEDULE\_COLLECTIVE} procedure (line 1) is called whenever a new collective is requested by the training workload. Lines 2--4 are for initialization. The \emph{for} loop in line 5 is for chunking the data while the lines 6--13 are executed to call the \texttt{scheduler} to determine the schedule of each chunk through \emph{SCHEDULER.SCHEDULE} procedure. Note that \sch assumes the \allgathershort schedule is the reverse order of obtained \reducescattershort schedule for a chunk if the collective is \allreduceshort (line 8). The \texttt{scheduler} first retrieves the current loads of all network dimensions via the \texttt{Dim Load Tracker} component (line 18). \texttt{Dim Load Tracker} is simply a list that contains the total load (chunk runtimes predicted by the \texttt{Latency Model}) that is placed on each dimension by the current schedules of the chunks. Lines 19--21 are for the robustness of \sch and check if the current load difference between the dimensions with maximum and minimum load is below a threshold. If this is true, then \sch reverts to the baseline scheduling to prevent oversubscribing the network dimensions with lower BW just because their current load is slightly lower than the dimensions with higher BW.   
If the condition of line 19 is false, \sch schedules the current chunk in a way to balance the loads across different dimensions. It first gets the index of dimensions sorted from least (most) load to most (least) load if the collective type is \reducescattershort (\allgathershort) (lines 21--27). This sorted list is the schedule for the new chunk, since such a schedule puts more load on the dimensions with currently lower loads, leading to filling the gap between the high-load and low-load dimensions. Next, the \texttt{Latency Model} predicts the load of the newly scheduled chunk (lines 28--29), and then \texttt{Dim Load Tracker} is updated (increased) accordingly (line 30). The \texttt{Latency Model} is a function that inputs chunk size, network dimension, and chunk operation (\reducescattershort/\allgathershort), and returns the predicted runtime for that chunk operation running on the specific dimension. Then, the scheduling process for the next chunk begins. 

\textbf{Example.} \autoref{fig:ThemisMechanism} shows how baseline vs. \sch scheduling works (i.e. assigns schedules) for the example of \autoref{fig:ThemisTimeDiagram}. As \autoref{fig:ThemisMechanism} shows, the baseline scheduling scheme always assigns a constant schedule for all chunks, hence, the gap between dim1 and dim2 preserves as new chunks are scheduled. However, \sch schedules the chunks differently to balance the dimension loads. In this example, \sch schedules the second chunk to start from dim2 to fill the gap between dim1 and dim2 (step c). After that, the last two chunks start from dim1 to fill the gap of dim1 with now overloaded dim2 (steps d\&e).
\autoref{fig:ThemisTimeDiagram}.b shows the \sch time diagram that is based on the schedule generated in \autoref{fig:ThemisMechanism}.b. As \autoref{fig:ThemisTimeDiagram}.b shows, such dimension load balancing results in better network utilization and reduced total communication time. We used a 2-dimensional example for simplicity. However, in general, the lack of ability to utilize the network BW in the baseline scheduling is more pronounced as the number of network dimensions and available extra BW of dimensions increase.

\subsection{Intra-Dimension Chunk Scheduling}\label{subsec:IntraDim}


So far, we have discussed how \sch schedules chunks across different dimensions to balance the loads (inter-dimension scheduling). Another question to answer is how different chunks within a dimension are ordered for processing because at any given point there might be multiple chunks available for each dimension. Adverse intra-dimension scheduling can lead to starvation of some dimensions since their, yet to come, chunks are stuck within the queues of other dimensions. 

We found out that in the baseline scheduling scheme, intra-dimension scheduling has minimal effects on the performance due to the identical schedule of all chunks. In other words, no matter how each dimension selects chunks to process, the average BW utilization remains fixed. The only difference is in the periods of time where the dimensions are utilized. Moreover, a monotonic schedule means each network dimension always receives the same chunk sizes. Hence, we assume the baseline scheme uses a simple FIFO-based intra-dimension chunk execution.

But for \sch, chunk intra-scheduling is important due to the different schedules of chunks, that result in variable chunk sizes per dimension. We empirically found the best policy to be Smallest-Chunk-First (SCF). The underlying intuition is that processing smaller chunks takes a shorter time and allows the chunk to be fed to other dimensions faster. This reduces the chance of a network dimension momentarily being idle due to not having collective chunks to process (i.e., dimension starvation).

Also note that if the chunk is small, processing one chunk per dimension underutilizes the network BW since small messages cannot saturate the given BW (e.g., due to the link latency). Hence, in this case, multiple chunks per dimension (if available) should be run in parallel to fully saturate a dimension's available BW (similar to the collective fusion concept in NCCL \cite{nccl}).

\subsection{Understanding All Latency Parameters}\label{subsec:AllLatency}
The techniques presented in \autoref{subsec:Algorithm} and \autoref{subsec:IntraDim} aim to \textbf{balance the total latency across different network dimensions}, since the collective performance is dictated by the slowest dimension. In general, the total latency of the K'th network dimension (dimK) can be calculated as follows:

\vskip -0.1in
\begin{equation*}
  \textit{Latency(dimK)} = A_{K} + (N_{K}\times B_{K}) +  idle_{K} 
\end{equation*}
\vskip 0.05in
While $A_{K}$ refers to the fixed delay caused by the collective algorithm and system latencies, $N_{K}$ is the total amount of bytes scheduled, $B_{K}$ is the per-byte latency, and $idle_{K}$ is the idle time, all corresponding to dimK. Among these parameters, $A_{K}$ and $B_{K}$ are given by the system specification and/or collective algorithm, while \sch controls $N_{K}$ and $idle_{K}$.

$A_{K}$ is the fixed delay to pay to run a certain collective type on a network dimension and is determined by:
\begin{equation*}
A_{K} = \text{number\_of\_steps} \times \text{step\_latency}
\end{equation*}
number\_of\_steps is determined by the basic collective communication algorithm (\autoref{subsec:BasicCollectiveAlgorithm}) employed on dimK. For example, ring-based \allreduce require $2P_{K}-2$ steps on dimK. On the other hand, \text{step\_latency} is determined by the network component latencies (e.g., NIC latency, link latency, etc.) when transferring a minimum-size message between two NPUs \cite{TACCL}. On real systems, $A_{K}$ can be calculated by running a minimum size collective on dimK.

To account for this delay in \sch, the \texttt{Dim Load Tracker} initializes each dimension's load to its respective $A_{K}$ for the target collective type (line 2 in Algorithm \autoref{alg:Themis}). Note that $A_{K}$ of different dimensions are considered to be negligible and not shown in the example of \autoref{fig:ThemisMechanism} for more readability.

The per-byte latency ($B_{K}$) is directly proportional to the inverse of link BW \cite{SCCL}, while $N_{K}$ corresponds to the total data size each NPU sends out on dimK and is calculated as follows: 

\vskip -0.03in
\[ N_{K} = \sum_{i=1}^{CPC} n_{K}^{i} \]
\vskip 0.03in

Where $n_{K}^{i}$ is the total amount of data each NPU sends out on dimK to process chunk \#i with respect to its schedule and collective algorithm \footnote{For example, if chunk \#i size is 4MB on dimK, then for ring-based RS/AG algorithm we have: $n_{K}^{i} = \frac{P_{K}-1}{P_{K}}\times \text{4MB}$.}.

Effectively, \sch (\autoref{subsec:Algorithm}) controls $N_{K}$, through dynamic scheduling of chunks,  to balance the overall latency across different dimensions. Since $N_{K}$ only participates with $B_{K}$, the \texttt{Latency Model} only considers $n_{K}^{i}\times B_{K}$ as the latency of chunk \#i on dimK (lines 28--29 in Algorithm \autoref{alg:Themis}).

The other factor is $idle_{k}$ that corresponds to the times where dimK network is idle, while there are other chunks stuck on other dimensions and yet have some pending stages on dimK to be executed later. To minimize the $idle_{k}$, we make 2 provisions as described in \autoref{subsec:IntraDim}. First, we employ SCF intra-dimension chunk scheduling to reduce the chance of dimension starvation. Second, if multiple available chunks are available, we execute multiple chunks per dimension if one chunk cannot fully saturate the network BW. 

\subsection{Supporting In-Network Collective Offload}\label{subsec:InNetworkCollective}
In recent years, several works have shown the communication performance improvement by offloading collectives to the switches belonging to different network dimensions~\cite{NVidiaSwitch,CommBottleneck1,NvidiaH100,FlareInNetwork}. Switch collective offload reduces the collective's network traffic (i.e., $n_{K}^{i}$) and fixed delay (i.e., $A_{K}$)~\cite{FlareInNetwork}. However, the concept of running hierarchical collectives, as described in \autoref{subsec:HierarchicalCollectives}, to cope with the heterogeneous multi-dimensional networks remains the same. Hierarchical collectives may suffer from the imbalanced network dimension loads as we discussed throughout this paper. Therefore, \sch can be used to balance the loads across different network dimensions even when in-network collective is available. 

\subsection{Chunk Schedule Consistency}
To design a distributed chunk scheduling algorithm, it is important that all NPUs execute the same order of chunks operations on their different dimensions. Failing to do so can lead to chunk schedule inconsistency and create a deadlock, since different NPUs wait on executing different chunk operations, and hence, no chunk can proceed \cite{nccl}. To maintain consistency, we must make sure that: (i) all NPUs produce the same chunk schedule (Inter-Dimension Schedule Consistency) and (ii) for a given dimension, all NPUs execute the same order of chunk operations (Intra-Dimension Schedule Consistency).

\subsubsection{Inter-Dimension Schedule Consistency} \label{subsubsec:intra_dim_cons}
Inter-dimension schedule consistency is guaranteed since both the \texttt{Latency Model} and \texttt{Dim Load Tracker} are similar across all NPUs, and behave in the same way as explained in \autoref{subsec:AllLatency}. This is possible because both $A_{K}$ and  $B_{K}$ parameters can be obtained offline and replicated across all NPUs. Therefore, different NPUs produce \textbf{exactly} the same schedule for the chunks of a collective operation. 

\begin{table*}[!t]
\centering
\caption{List of target topologies and their BW/latency configurations per each dimension. The naming convention starts with the number of dimensions followed by dimension topology time in increasing order. For example, \FcRingSwitch means a 3-dim topology where dim1 is FullyConnected, dim2 is Ring, and dim3 is Switch. Each color shows a dimension in which the network connects the NPUs together.
The BW and latency in each dimension are selected according to the predicted ranges for link technologies in the future systems, listed below the table. We note that based on technology trends, there is a wide range of BW and latency options for each dimension depending on the technology and other constraints. All combinations cannot be presented here due to the lack of space. We create a diverse set of topologies with a different BW ratio to motivate the problem and demonstrate the applicability of Themis on various platforms. The Aggr BW/NPU is the product of the BW/link determined by the technology and Links/NPU is determined by the topology.
} 
\vskip -0.1in
\resizebox{\textwidth}{!}{
\begin{tabular}{|l|c|c|c|c|c|c|}
\hline
\multicolumn{1}{|c|}{\textbf{Name}} & \textbf{NPUs} & \textbf{Size} & \textbf{BW/Link (Gb/s)**} & \textbf{\#Links/NPU} & \textbf{Aggr BW/NPU (Gb/s)} & \textbf{Network Latency (ns)**} \\ \hline
\SwitchSwitchColored & 1024 & \teal{16}$\times$\blue{64} & (\teal{200}, \blue{800}) & (\teal{6}, \blue{1}) & (\teal{1200}, \blue{800}) & (\teal{700}, \blue{1700}) \\ \hline
\switchhomoColored & 1024 & \teal{16}$\times$\blue{8}$\times$\violet{8} & (\teal{200}, \blue{200}, \violet{800}) & (\teal{4}, \blue{4}, \violet{1}) & (\teal{800}, \blue{800}, \violet{800}) & (\teal{700}, \blue{700}, \violet{1700}) \\ \hline
\switchheteroColored & 1024 & \teal{16}$\times$\blue{8}$\times$\violet{8} & (\teal{200}, \blue{200}, \violet{400}) & (\teal{8}, \blue{4}, \violet{1}) & (\teal{1600}, \blue{800}, \violet{400}) & 
(\teal{700}, \blue{700}, \violet{1700}) \\ \hline
\FcRingSwitchColored & 1024 & \teal{8}$\times$\blue{16}$\times$\violet{8} & (\teal{200}, \blue{200}, \violet{400}) & (\teal{7}, \blue{4}, \violet{1}) & (\teal{1400}, \blue{800}, \violet{400}) & (\teal{700}, \blue{700}, \violet{1700}) \\ \hline
\RingSwitchSwitchSwitchColored & 1024 & \red{4}$\times$\teal{4}$\times$\blue{8}$\times$\violet{8} & (\red{1000}, \teal{200}, \blue{200}, \violet{400}) & (\red{2}, \teal{8}, \blue{4}, \violet{1}) & (\red{2000}, \teal{1600}, \blue{800}, \violet{400}) & (\red{20}, \teal{700}, \blue{700}, \violet{1700}) \\ \hline
\RingFcRingSwitchColored& 1024 & \red{4}$\times$\teal{8}$\times$\blue{4}$\times$\violet{8} & (\red{1500}, \teal{200}, \blue{200}, \violet{800}) & (\red{2}, \teal{7}, \blue{6}, \violet{1}) & (\red{3000}, \teal{1400}, \blue{1200}, \violet{800}) & (\red{20}, \teal{700}, \blue{700}, \violet{1700}) \\ \hline
\end{tabular}
}
\label{table:topology_description}
\begin{flushleft}
\footnotesize{
**Link Technologies for each Dimension: \textcolor{red}{chiplet-to-chiplet (within a package)}
~\cite{mcm-gpu,Simba,AMDInfinity,saeedACE}, \textcolor{teal}{package-to-package (within a server node)}~\cite{NVLink3,IntelExaScale,HabanaPtP,TACCL}, \textcolor{blue}{node-to-node}~\cite{HabanaPtP,NIC400G,NIC800G,TACCL}, \textcolor{violet}{pod-to-pod}.}~\cite{NIC400G,NIC800G,CiscoLatency,TACCL}. Note - for all topologies, the last dimension uses NICs, which is node-to-node for 2D, and pod-to-pod for 3D and 4D. The network latency (i.e., step\_latency in \autoref{subsec:AllLatency}) corresponds to the direct NPU-to-NPU latency when sending a minimum-length message.
\end{flushleft}
\end{table*}
\subsubsection{Intra-Dimension Schedule Consistency} Runtime variation might rarely result in chunks being available for a given dimension in different orders across different NPUs. For example, suppose the chunk operations C1.1 and C2.1\footnote{The notation is based on \autoref{fig:ThemisTimeDiagram}.} are under execution on dim1 and dim3, respectively, and their next operation (i.e., C1.2, and C2.2) is scheduled on dim2. Runtime effects (e.g., packet drop, endpoint congestion) might cause C1.1 to slightly finish sooner than others, followed by the immediate start of C1.2 on dim2 on some NPUs. Now suppose on some NPUs with unfinished C1.1, C2.1 is finished sooner, and hence, those NPUs begin running C2.2 on dim2, creating a potential deadlock case.

To prevent this, once Inter-Dimension Schedules are determined (according to \autoref{subsubsec:intra_dim_cons}), \sch simulates their execution to get an estimation of when each chunk operation will be available on each dimension. Once chunk operation availability is estimated, \sch enforces this intra-dimension ordering on runtime. Even if some chunks are available sooner on the NPU, \sch does not execute them if it is not their turn to be executed.

Note that the simulation is deterministic, so all NPUs produce the same intra-dimension ordering. In addition, the simulation does not need to consider detailed network modeling and is performed fast, since the aim is the order of chunk availability on each dimension, and not their exact availability time. Once a certain collective schedule and its ordering are generated by \sch, it is saved and reused on later training iterations. So, there is no need to repeat the process in later training iterations.

\begin{table}[]
\centering
\caption{Target Collective Schedulers}
\label{table:Configs}
\vskip -0.14in
\resizebox{\linewidth}{!}{%
\begin{tabular}{|c|l|}
\hline
\multicolumn{1}{|c|}{\textbf{Method}} & \multicolumn{1}{c|}{\textbf{Comment}} \\ \hline
Baseline & \begin{tabular}[c]{@{}l@{}}Uses multi-rail hierarchical algorithm \cite{blueconnect} as explained\\in \autoref{subsec:HierarchicalCollectives} with FIFO intra dimension scheduling.\end{tabular}                        \\ \hline
Themis+FIFO & Uses \sch with FIFO intra-dimension scheduling. \\ \hline
Themis+SCF  & Uses \sch with SCF intra-dimension scheduling.  \\ \hline
\ideal    & \begin{tabular}[c]{@{}l@{}}Assumes 100\% BW is utilized. Communication latency\\is simply calculated by (collective size / total BW).\end{tabular} \\ \hline
\end{tabular}%
}
\end{table}
\setlength{\textfloatsep}{4mm}
\section{Methodology}\label{sec:methodology}

In this section, we provide our methodology and the target systems and workloads to evaluate \sch and baseline.

\subsection{Simulation Platform}
\label{sec:simulation}
We use the ASTRA-SIM simulator \cite{AstraSimGithub,astrasim} to implement our scheme and compare it with the baseline system. ASTRA-SIM provides the flexibility to define various large-scale hierarchical training platforms, enabling us to demonstrate the efficiency of \sch on future platforms. ASTRA-SIM simulates the communication performance of the distributed training workloads in detail. It supports heterogeneous networks, different collective communication algorithms, and different parallelization strategies. 
For compute times (in the case of real workloads) we assumed roofline FP16 performance from the total FLOPS available on current state-of-the-art accelerators~\cite{A100}.

We note that, since we are targeting next-gen systems, using simulation as our evaluation methodology is our only option to show the necessity of \sch for such systems. Recall that regular topologies and hierarchical topology-aware collective communication algorithms (\autoref{subsec:BasicCollectiveAlgorithm}) per dimension lead to congestion-less network traffic, enabling detailed simulators to accurately model and match real system measurements for collectives~\cite{AstraSimGithub,astrasim}.

\vspace{-1.9mm}
\subsection{Training Platforms and Workloads}\label{subsec:TargetPlatforms}
\autoref{table:topology_description} shows our  target topologies all consisting of 1024 NPUs to resemble large-scale next-generation systems.
\autoref{subsec:modelingSystems} presents more description of trends and previous works that leads to the topologies presented in \autoref{table:topology_description}. We do not consider in-network collective offload support due to the lack of space, although we expect \sch still improves the network BW utilization in this case, as explained in \autoref{subsec:InNetworkCollective}.

\betterparagraph{All-Reduce Algorithm} 
\autoref{table:TopologyAwareCollectives} lists the topology-aware and contention-free AR algorithms we employ.





\betterparagraph{Target Workloads and Parallelization}
\sch is a solution to maximize the BW utilization of pervasive collective communications on multi-dimensional networks. Collective communication is an integral part of any synchronous training job with an NPU-to-NPU communication mechanism. Moreover, collective usage is not limited to training only, and has been widely used in other domains (e.g., in HPC applications and distributed inference). However, we limit the scope of our evaluations in this work to DL training given its importance.

For real workload training, we selected four DNNs from different domains of deep learning applications: \resnet \cite{ResNet} (from computer vision), \dlrm \cite{DLRM} (from recommendation models), \gnmt \cite{gnmt}, and \transformerlarge (one trillion parameter) \cite{Transformer1T} (both from NLP domain). For \dlrm, we use the model described in \cite{hotiPaper}. The gradient precision is FP16 in all workloads and the per-NPU minibatch size is set to be 32, 512, 128, and 16 for \resnet, \dlrm, \gnmt, and \transformerlarge, respectively. Such workloads have a high ratio of communication-to-computation and hence, benefit most from applying \sch.

In terms of parallelization strategy, \resnet and \gnmt use the complete data-parallel partitioning since they can fit within single NPU's memory. \dlrm uses data-parallel partitioning for its MLP layers, while its sparse features (embedding tables) are partitioned in model-parallel. To reduce the memory requirements for DNN training, \transformerlarge uses Microsoft ZeRO optimizer stage 2 \cite{mszero}. \transformerlarge is partitioned in a model-parallel manner across the first dimensions up to 128 NPUs, and data-parallel across the remaining dimensions. The reason is that a single NPU memory is usually within the range of 48--64GB \cite{HabanaPtP,IntelExaScale}. Thus, the entire parameters of \transformerlarge (even after applying ZeRO optimizer) can not fit on a single NPU, requiring model-parallel to split the model.

\betterparagraph{Multi-Tenancy}
We target systems that are private training clusters dedicated to training a DNN workload at a time without other interfering workloads\footnote{In fact, many previous collectives algorithm works have assumed the same environment (e.g., \cite{SCCL,BLink}).}. Therefore, the NPU network only observes a single workload training traffic. Such platforms are common enough to be considered separately and are widely deployed in the industry to train critical workloads \cite{FBDLRMPlatform,googleCloudTpu,NVIDIASuperPod}.

\subsection{Target Configurations}

\betterparagraph{Target Scheduling Configurations}
\autoref{table:Configs} shows the scheduling policies we implement. The baseline uses FIFO intra-dimension policy as different intra-dimension policies have no effect on its performance (discussed in \autoref{subsec:IntraDim}).

To decompose the effect of inter-dimension and intra-dimension scheduling, we present two flavors of \sch: i) \schfifo that uses the default FIFO intra-dimension scheduling policy, and ii) \schscffull (SCF) with optimized SCF intra-dimension policy. Moreover, for the real workloads, we implement an \ideal method that assumes 100\% network BW utilization. The ideal method determines the upper bound for maximum achievable speed-up and guarantees that no chunk scheduling scheme can exceed its performance.

\betterparagraph{\sch Parameter Values}
According to \autoref{fig:ThemisStructure} and Algorithm \autoref{alg:Themis}, \sch has two important parameters to be set: the number of chunks per collective and the Threshold (line 19 in Algorithm \autoref{alg:Themis}). Unless mentioned otherwise, we set the number of chunks per collective to be 64 in all our experiments for both the baseline and \sch. We set the Threshold to be the estimated runtime (predicted by the \texttt{Latency Model}) when running an RS/AG of size  $\frac{\text{chunkSize}}{16}$ on the dimension with the lowest current load.

\section{Results}\label{sec:results}


In \autoref{subsec:MicrobenchmarkResults}, we first present the single collective microbenchmark results and dive deep into the reasons for \sch showing benefits over the baseline scheduling scheme. Next, in \autoref{subsec:RealWorkloadResults} we present the end-to-end training iteration results for real workloads such as \resnet, \gnmt, \dlrm, and \transformerlarge. Finally, in \autoref{subsec:InsightsForFuture}, we give insights on how next generation  networks should be designed in terms of BW distribution for distributed training. 

\subsection{Microbenchmark Results}\label{subsec:MicrobenchmarkResults}
\vspace{-1mm}
\autoref{fig:GraphCommstimeCommscaleUpdated} shows the \allreduce communication results of the baseline and \sch, ranging from 100 MB to 1 GB. We pick this range to represent the (relatively) large models' collectives that are the target of such large-scale distributed training systems. This range also covers our target workloads collectives in  \autoref{subsec:RealWorkloadResults}. 

As \autoref{fig:GraphCommstimeCommscaleUpdated} shows, applying \sch significantly reduces the communication time. When averaging across all topologies and comm sizes, \schfifo and \schscf reduce the communication time by \WW{1.58$\times$ and 1.72$\times$} over the baseline, respectively.

\begin{figure*}[ht]
\vspace{2mm}
\centering
    \begin{minipage}[b]{0.63\textwidth}
        \includegraphics[width=0.99\linewidth]{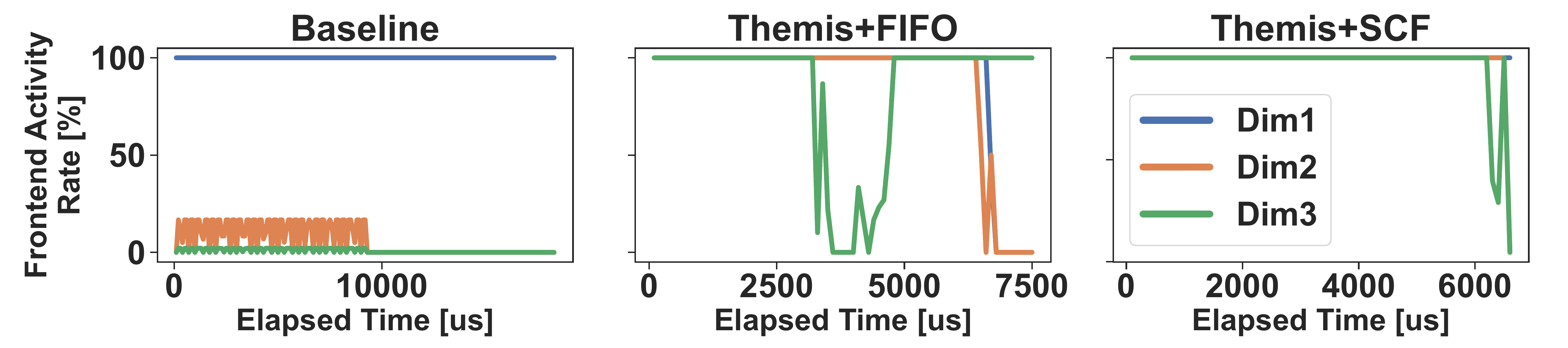}
        \vspace{-4mm}
        \caption{The average per-dimension activity rate for, \schfifo, and \schscffull (SCF) for a 1GB \allreduce size when running on \switchhomo topology. A dimension is called to have activity if there is at least one chunk in that dimension for processing at any given point in time. Frontend activity rate is obtained by calculating the percentage of times each dimension has activity during a period of 100$\mu s$.}
        \label{fig:GraphFrontendActivityTimeUpdated}
    \end{minipage}
    \quad
    \begin{minipage}[b]{0.34\textwidth}
        \includegraphics[width=0.99\linewidth]{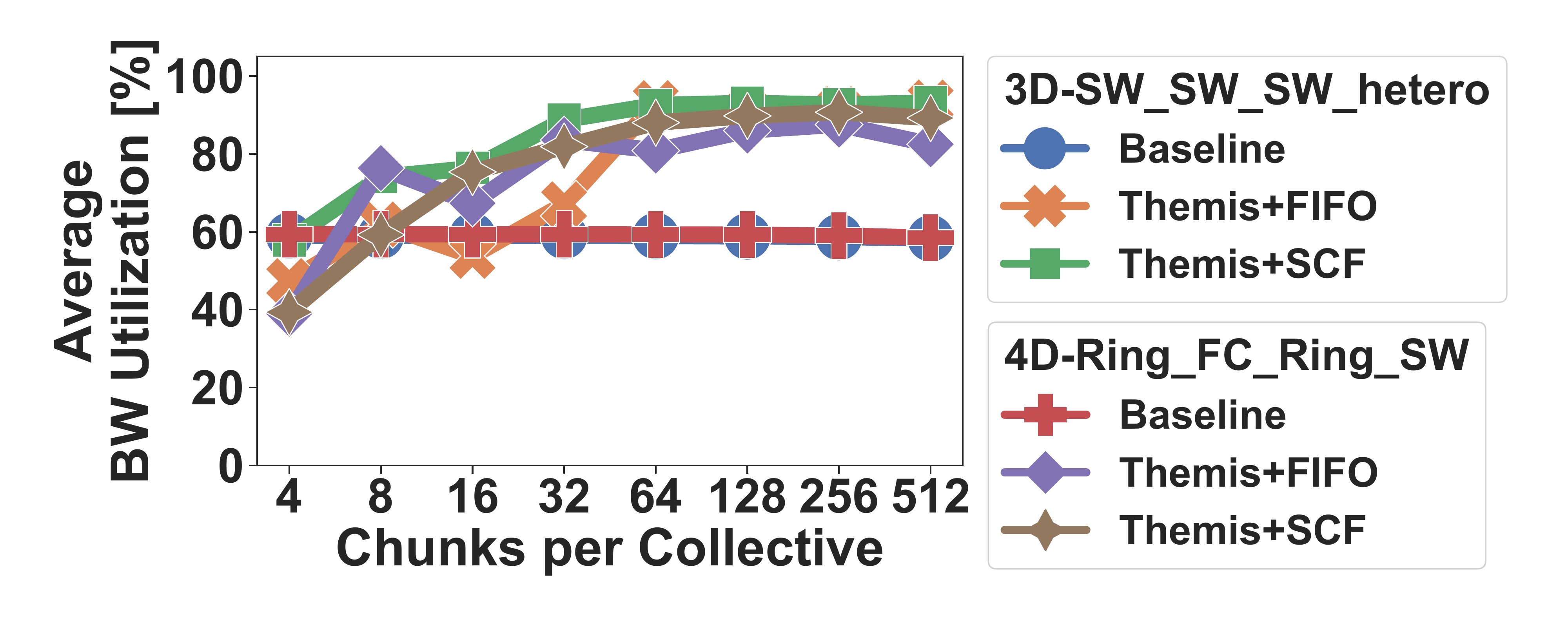}
        \vspace{-4mm}
        \caption{Sensitivity analysis that shows total BW utilization for a single 100MB All-Reduce when the number of chunks per collective varies from 4 to 512. The target topologies are \switchhetero and \WW{\RingFcRingSwitch}.}
        \label{fig:GraphSensetivity}
    \end{minipage}
\vspace{1mm}
\end{figure*}

To shed light on the reason behind \sch benefits, \autoref{fig:GraphFrontendActivityTimeUpdated} shows the per-dimension frontend activity rates for a 1GB \allreduce on \switchhomo. A network dimension is called to have activity if there is at least one chunk in that dimension for processing at any given point in time. As can be seen, in the baseline system dim2 and dim3 show significant underutilization. The reason is dim1 is the bottleneck stage in the baseline pipeline scheduling and the unbalanced stage latencies result in underutilization. Both \schfifo and \schscf significantly balance the loads and improve the utilization of dim2 and dim3. An interesting point about \schfifo is the occasional underutilization of different dimensions. This is due to the inefficient FIFO intra-dimension chunk processing that leads to the starvation of some chunks as discussed in \autoref{subsec:IntraDim}. \schscf further reduces the starvation problem as can be seen in \autoref{fig:GraphFrontendActivityTimeUpdated}.

As \autoref{fig:GraphCommstimeCommscaleUpdated} suggests, the amount of speed-up obtained by \sch varies by the topology. The speed-up depends on the amount of underutilization in the baseline scheduling. For example, in the case of \switchhomo, and according to the discussion in \autoref{sec:motivation}, the baseline was able to achieve near-optimal performance if:
$$\text{BW(dim1)}=\text{16(dim2)}=\text{128BW(dim3)}$$
According to the \autoref{fig:GraphFrontendActivityTimeUpdated}, dim1 is the bottleneck. Therefore, in the case of \switchhomo, if we substitute BW(dim1) with 800Gbps, then we have:
$$800\text{Gbps}_{\text{dim}1}=16\times50\text{Gbps}_{\text{dim}2}=128\times6.25\text{Gbps}_{\text{dim}3}$$

Hence, 750Gbps of dim2 and 793.75Gbps of dim3 are underutilized by the baseline scheduling for \switchhomo.

\insertWideFigureNewSpaceBottom{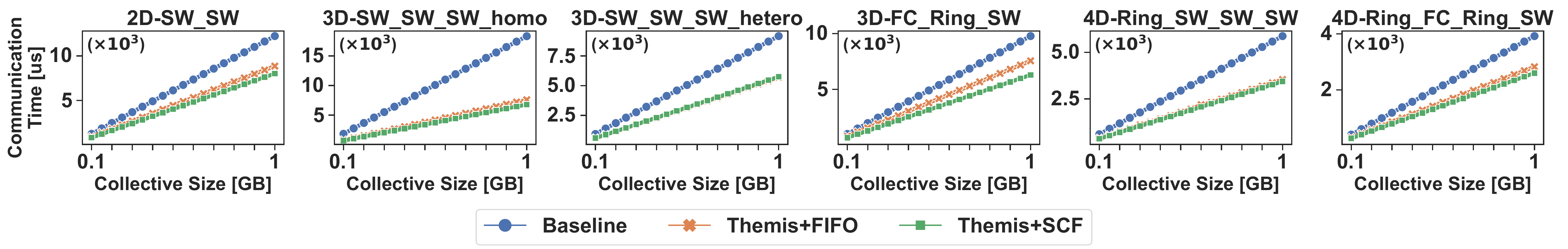}{Total communication time of baseline, \schfifo, and \schscf for different size All-Reduces. Note that \schfifo uses FIFO intra-dimension chunk scheduling, while \schscf uses the smallest available chunks for intra-dimension chunk execution.}{1}{-5}{-3}

\insertWideFigureNewSpaceBottom{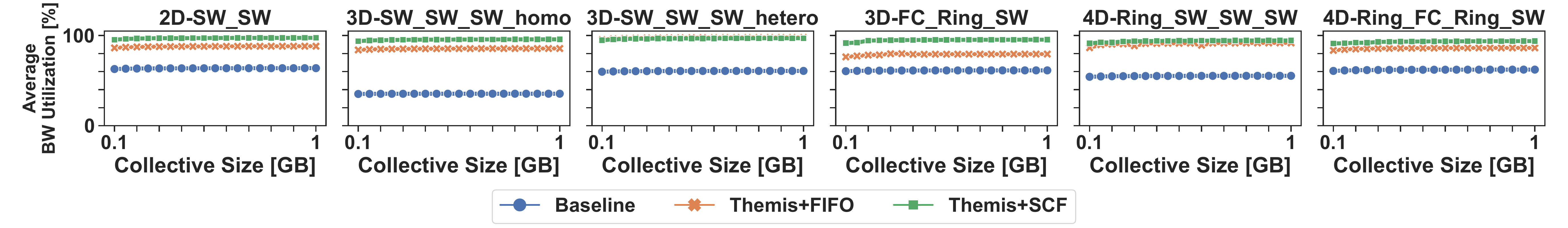}{Average BW utilization of baseline, \schfifo, and \schscffull (SCF) for different size All-Reduces.}{1}{-5}{-2}

\autoref{fig:GraphBWUtilizationCommscaleUpdated} shows the average network utilization for variable sizes \allreduce sizes. In general, as the collective size increases, its performance becomes more BW bound and the network latency component is minimized, leading to increased BW utilization. But the baseline scheduling cannot saturate the full BW as a result of the fundamental mismatch between different stage latencies of the hierarchical collective algorithms. On average, baseline, \schfifo, and \schscf can achieve \WW{56.31\%, 87.67\%, and 95.14\%} of the network BW utilization, respectively. This indicates that \sch is an efficient method that can exploit and leverage almost all underutilized opportunities which exist in the baseline, leaving less room for further optimizations.

Next, we study the effect of chunk granularity on the performance of \sch. \autoref{fig:GraphSensetivity} shows the BW utilization for different number chunk granularities for baseline and \sch when running on \switchhetero and \WW{\RingFcRingSwitch} topologies. Other topologies are not included due to space limitations. For the baseline, dim1 is the bottleneck (on both topologies) and the latency is mostly determined by the rate dim1 receives the chunks to process. Since dim1 is always the first dimension to receive the chunks in the baseline scheduling, changing the chunk granularity does not significantly affect its performance. However, increasing the number of chunks (decreasing chunk size) enables \sch to better balance the loads across the dimensions. When increasing the chunks from 4 to 512, BW utilization for \schscf (\schfifo) increases from \WW{48.58\% (43.13\%) to 91.18\% (87.81\%)} on average across the two topologies.
In contrast, increasing the number of chunks per collective (reducing chunk size) might eventually reduce the individual chunk network operations to go below the max packet size on some network dimensions. This increases the header-to-packet ratio and hurts the network's \textit{goodput}.

We picked the default number of chunks to be 64 that achieve \WW{95.14\%} BW utilization for the microbenchmark workload when averaging across all of our target topologies and collective sizes. This comes at the expense of increasing the total header-to-packet ratio by less than $0.5\%$ in the worst case (i.e., 100 MB \allreduceshort), when compared to 1 chunk per collective for the microbenchmark workload\footnote{We assume the chiplet-to-chiplet packet format is similar to \cite{NOCPacketFormat}, rack-scale format is similar to \cite{NVLink3}, and the NIC packet format is  InfiniBand RC mode with the remote write verb \cite{InfiniBandPacketFormat}.}.  

As can be seen in \autoref{fig:GraphSensetivity}, at some points, increasing chunks modestly reduces the BW utilization for \sch. This is mainly because of the starvation case discussed earlier. However, \schscf shows stable behavior starting from \WW{8} chunks on all of our tested topologies.

\insertWideFigureNewSpace{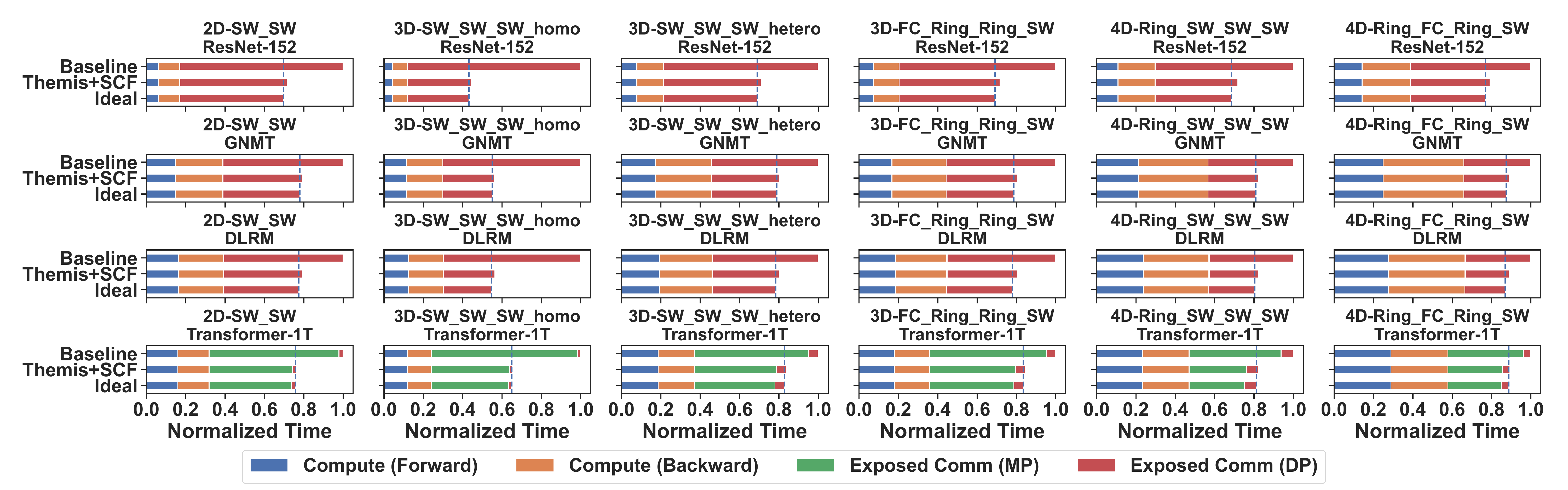}{Training times for 3 iterations for \resnet, \gnmt, \dlrm, and  \transformerlarge running on different topologies. Training iteration consists of a forward-pass followed by a back-propagation step. The total latency is decomposed to the compute times (across all layers), plus the total exposed communication latency. Compute times stem from computation during the forward pass (blue bar) or during the backpropagation (orange bar). Exposed communication may be due to the waiting for the data-parallel communications (red bar), or model-parallel communication (green bar), as explained in \autoref{subsec:TargetPlatforms}. For each workload, the latency of the baseline is normalized to 1.}{1}{-5}

\subsection{Real Workload Results}\label{subsec:RealWorkloadResults}
\vspace{-1mm}
In this section, we present real workload results to find the effect of \sch on the total end-to-end training iteration times which we break into \emph{total computation} + \emph{exposed communication}.\footnote{Exposed communication refers to the communication overhead of the training time where the training workload is waiting for the communication to be finished.}. Here, we only use \schscf configuration since it was shown to be the better approach in \autoref{subsec:MicrobenchmarkResults}.

In our case and for the data-parallel partitioning, exposed communication occurs at the end of back-propagation, where NPUs communicate their locally computed weight gradients through \allreduce, updating their model parameters before the next iteration starts.

Handling the model-parallel communication case is different in \dlrm vs. \transformerlarge. For DLRM, its sparse features form a concurrent path with bottom-MLP layers, and therefore, its model-parallel communication (in terms of All-to-All collective operation) is performed in parallel with forward-pass, and back-propagation of bottom-MLPs. We only wait for the embedding communication operation (i.e. all-to-all) before entering the top-MLP layers in forward-pass, and after finishing the back-propagation to update the embedding. In the case of \transformerlarge, the output-activations/input-gradients of a (model-parallel) layer must be communicated (through \allreduceshort or \allgathershort, depending on the layer type in Transformer) during forward-pass/back-propagation before processing the next layer. \autoref{fig:GraphWorkloadRuntimeUpdated} shows the training iteration times that are decomposed into total compute time and total exposed communication time.

For training, back-propagation computation usually takes longer since it needs to compute for both weight gradients and input gradients, compared to the forward-pass that only involves forward computation. However, this is not the case for \transformerlarge since it consists of forward-in-back-propagation steps, as a result of ZeRO optimizer, that is counted towards forward-pass in \autoref{fig:GraphWorkloadRuntimeUpdated}.

As \autoref{fig:GraphWorkloadRuntimeUpdated} shows, \resnet and \gnmt only experience data-parallel exposed communication since they are distributed in pure data-parallel. An interesting point is about \dlrm where it has a hybrid (data+model parallel as explained in \autoref{subsec:TargetPlatforms}) parallelism, but only the data-parallel communication is counted towards exposed communication. This is because model-parallel communication (All-To-All) is overlapped with the forward-pass, back-propagation operations of bottom-MLP layers. In \transformerlarge the model-parallel communication is the dominant factor. Also, note that the data-parallel communication of \transformerlarge uses only the last network dimension in all of the topologies. This indicates that there is only one scheduling possible for data-parallel communication of \transformerlarge, meaning that baseline and \sch have the same performance for this portion of the exposed communication. When averaging across all topologies and workloads, applying \sch reduces the exposed communication time by \WW{1.65$\times$}. The speedup is close to the \ideal system which reduces the exposed communication time by \WW{1.72$\times$}, on average. 

Such reduction in exposed communication leads to a reduction in overall training time as well. However, the overall training iteration benefit follows Amdahl's law \cite{AmdahlsLaw} and depends on the current ratio between the exposed communication and total computation. When averaging across all topologies, \sch reduces the training iteration time by \WW{1.49$\times$ (2.25$\times$ max), 1.30$\times$ (1.78$\times$ max), 1.30$\times$ (1.77$\times$ max), and 1.25$\times$ (1.53$\times$ max)} for \resnet, \gnmt, \dlrm, and \transformerlarge, respectively. On the other hand, the \ideal system achieves training iteration speed-up of \WW{1.54$\times$, 1.32$\times$, 1.33$\times$, and 1.26$\times$} for \resnet, \gnmt, \dlrm, and \transformerlarge, respectively. \textit{Overall, we find \sch is close to the ideal system, leaving little opportunity for further optimization.}

\subsection{Insights for Future System Design}\label{subsec:InsightsForFuture}
\vspace{-1mm}
Throughout this paper, we showed how \sch can drive the BW of all network dimensions. This raises the question that how architects and system engineers should distribute the network BW across different network dimensions in the first place and whether some design points should be prohibited since even Themis cannot help. Consider any two dimensions dimK and dimL of the network, where K<L and $P_{I}$ to be the network size in dimI. In this section, we describe three different scenarios for BW distribution, depending on the BW provision for dimL: 

\textbf{Just Enough BW Scenario.} Here, the baseline (and \sch) scheduling algorithm can fully utilize the network. As explained in \autoref{sec:motivation}, the BW distribution should be:
$$\text{BW(dimK)}=P_{K}\times P_{K+1}\times ....\times P_{L-1}\times \text{BW(dimL)}$$
In this case, the chunk size ratio is proportional to the BW ratio of the two dimensions. Hence, the baseline algorithm is sufficient to utilize both dimensions.

\textbf{OverProvisioned BW Scenario.} 
$$\text{BW(dimK)}<P_{K}\times P_{K+1}\times ....\times P_{L-1}\times \text{BW(dimL)}$$

As explained in \autoref{sec:motivation}, this is the case where the baseline can not utilize the full BW of dimL. While \sch redistributes the loads that result in full utilization of both dimensions.

\textbf{UnderProvisioned BW Scenario.} In this case there may be no scheduling algorithm that can fully drive both dimensions:
$$\text{BW(dimK)}>P_{K}\times P_{K+1}\times ....\times P_{L-1}\times \text{BW(dimL)}$$
In such BW distribution and with baseline scheduling, dimK is underutilized while it has $P_{K}\times P_{K+1}\times ....\times P_{L-1}$ more loads, compared to dimL (dimL has underprovisioned BW). 

To fully utilize both dimensions, any redistribution of chunks should increase the load of dimK compared to dimL. However, this only can happen if dimK has overprovisioned BW compared to some other dimension and this might not always be the case. For example, in a simple 2-dimensional network case where K=1, L=2, there is no scheduling that can fully utilize both dimensions, since the baseline scheduling already puts the highest load on dimK, and any other scheduling increases the load gap between dimK and dimL (rather than reducing it). Thus, such design points should be prohibited.

\vspace{-1mm}
\section{Related Works}
\textbf{HPC Platforms.} Collective communication algorithms are vastly studied in the context of High Performance Computing (HPC) workloads using the MPI communication interface \cite{MPI}. Many different implementations of MPI interface are proposed in \cite{collective1,collective10,collective11}, as well as topology-aware algorithms \cite{collective12,collective13,HierarchicalMPICollective}, and efficient collective execution on shared-memory processor clusters \cite{collective15,collective16}. 
Nonetheless, these CPU-based collective algorithms either assume BW-symmetric topology or apply hierarchical algorithms with a \textbf{fixed schedule}. Moreover, collectives in HPC are usually small (few KB-MB) unlike DL (10s of MB-GB) where they lie in the critical path~\cite{NVidiaSwitch}.



\textbf{DL Training Platforms (1-dimensional).} Recently, collective communications are revisited for direct NPU-to-NPU communications for distributed DNN workloads. SCCL~\cite{SCCL} provides topology-aware pareto-optimal collectives for a given topology. EFLOPS \cite{EFLOPS} proposes a BiGraph topology with optimized HDRM collective algorithm. However, these works are based on BW-symmetric and 1-dimensional networks only.

\textbf{DL Training Platforms (Multi-dimensional).} In addition, collective communication libraries such as \cite{nccl,oneccl,GC3} provide a suite of collective algorithms optimized for various topology and collective sizes. Authors in \cite{BLink,PLink,blueconnect} take into account the physical topology hierarchies to perform localized reduction/aggregation per each network hierarchy. Blink \cite{BLink} is a framework to generate efficient collective algorithms based on the underlying network resources using the concept of packing spanning trees. PLink \cite{PLink} is a collective scheme that aims to cope with the heterogeneous network and variable performance of public cloud platforms. It creates a logical 2-dimensional network based on the distance proximity of VMs, and then applies a hierarchical collective algorithm to optimize for heterogeneous cloud network, and dynamically changes the collective algorithm within each (logical) network dimension to adapt to the performance variability due to other interfering workloads. However, all of these works perform hierarchical collectives with the \textbf{static schedule} of chunk operations across different network dimensions, which is not efficient for the next-gen platforms as discussed in this paper.

In contrast, \sch is the first method that proposes a \textbf{\textit{dynamic}} chunk scheduling for maximum BW utilization. Themis is also orthogonal to all previous hierarchical collective methods, meaning that it can leverage any proposed collective algorithm for each dimension, while only changing the schedules of each individual chunk.

\vspace{-1mm}
\section{Conclusion}\label{sec:conclusion}
In this paper, we identified
that hierarchical multi-stage collective algorithms fail to saturate the network BW of next-gen platforms due to different pipeline stage latencies induced by different chunk sizes and network characteristics in each network dimension. We proposed \sch as a solution to improve the BW utilization by dynamically scheduling the chunks to balance the loads across different network dimensions.

\sch improves the end-to-end training iteration performance of real workloads, such as \resnet, \gnmt, \dlrm, and \transformerlarge, by \WW{1.49$\times$ (2.25$\times$ max), 1.30$\times$ (1.78$\times$ max), 1.30$\times$ (1.77$\times$ max), and 1.25$\times$ (1.53$\times$ max)}, respectively.

\begin{acks}
This work was supported by awards from Intel and Meta. We would like to thank the reviewers for their insightful comments. We also thank Taekyung Heo for his help in revising the paper.  
\end{acks}

\bibliographystyle{ACM-Reference-Format}
\bibliography{bib/refs}


\end{document}